\documentclass[symmetry,review,accept,pdftex,moreauthors]{Definitions/mdpi} 
\firstpage{1} 
\pubvolume{1}
\issuenum{1}
\articlenumber{0}
\pubyear{2025}
\copyrightyear{2025}
\externaleditor{name}
\datereceived{31 January 2025} 
\daterevised{26 February 2025} % Comment out if no revised date
\dateaccepted{ } 
\datepublished{ } 
\hreflink{https://\\doi.org/} % If needed use \linebreak
\usepackage{amsfonts}
\usepackage{slashed}
\usepackage{bm}
\usepackage{verbatim} %comment
\usepackage{xcolor}
\makeatletter
\let\c@lofdepth\relax
\let\c@lotdepth\relax
\makeatother
\usepackage{subfigure}
\makeatletter
\renewcommand{\@thesubfigure}{\normalsize(\textbf{\alph{subfigure}})}
\makeatother

\def\be{\begin{eqnarray}}
	\def\ee{\end{eqnarray}}

\Title{Baryon Construction with $\eta^\prime$ Meson Field}
\TitleCitation{Baryon  Construction with $\eta^\prime$ Meson Field}

\Author{Fan Lin $^{1,2,3}$ and Yong-liang Ma $^{4,5,}$*}

\AuthorNames{Fan Lin and Yong-liang Ma}

\AuthorCitation{Lin, F.; Ma, Y.-l.}
\address{%
$^{1}$ \quad School of Fundamental Physics and Mathematical Sciences, Hangzhou Institute for Advanced Study,  UCAS, %MDPI: please add the full title if possible. same for aff.5.
 Hangzhou 310024,  China; linfan19@mails.ucas.ac.cn\\ %MDPI: We added the email addresses here according to those submitted online at susy.mdpi.com. Please confirm.
$^{2}$ \quad Institute of Theoretical Physics, Chinese Academy of Sciences, Beijing 100190,  China\\
$^{3}$ \quad  University of Chinese Academy of Sciences, Beijing 100049,  China\\
$^{4}$ \quad School of Frontier Sciences, Nanjing University, Suzhou 215163, China\\
$^{5}$ \quad International Center for Theoretical Physics Asia-Pacific (ICTP-AP), UCAS, Beijing 100190, China
}

\corres{Correspondence: ylma@nju.edu.cn}

\abstract{In the low-energy regime, baryons with $N_f \geq 2$ have long been constructed as skyrmions or through bag models, but such constructions for $N_f = 1$ are hindered by the trivial topological structure of the meson field. Recent proposals suggest that one-flavor baryons can instead be interpreted as quantum Hall droplets on the $\eta'$ domain wall, providing a potential link to quark--hadron continuity at high density. In retrospect, the qualitative or semi-qualitative construction of one-flavor baryons on the $\eta'$ domain wall reveals that these baryons can be described as quantum Hall droplets, resembling topological solitons akin to skyrmions. Using an effective theory on the $\eta'$ domain wall, which is conjectured to be the Chern--Simons--Higgs theory, it is discussed that its vortex solution with unit baryon numbers naturally has a spin of $N_c/2$, and thus can be interpreted as a baryon or multi-baryon structure. The particle--vortex duality suggests that quarks carry a fractional topological charge of $1/N_c$ and obey fractional statistics. In terms of chiral bag models, confinement can be attributed to the monopoles confined within the bag, and the vector meson fields on the bag surface are essential for ensuring the correct baryon number in the chiral bag framework, thereby providing deeper insights into baryons as non-trivial topological structures of the meson field. In this paper, we review the progress in this development, with a special focus on the $\eta^\prime$ domain wall dynamics. Naive extensions to $N_f \geq 2$ are also discussed.
}


\begin{document}

%%%%%%%%%%%%%%%%%%%%%%%%%%%%%%%%%%%%%%%%%%

\section{Introduction}

At high energies, quantum chromodynamics (QCD) is well described by a weakly interacting $SU(N_c)$ gauge theory coupled to $N_f$ fundamental fermions, where perturbative techniques are applicable~\cite{Politzer:1973fx,Gross:1973id,Marciano:1977su}. In contrast, at low energies, QCD transitions into a strongly coupled regime, leading to the confinement of quarks and gluons into color-neutral hadrons such as mesons and baryons. The lightest pseudoscalar mesons, identified as Nambu--Goldstone bosons, emerge from the spontaneous breaking of chiral symmetry: $SU(N_f)_L \times SU(N_f)_R \to SU(N_f)_V$. These mesons govern the dynamics of QCD at low energies, giving rise to a non-linear sigma model that serves as the low-energy effective theory for pions~\cite{Coleman:1969sm,Callan:1969sn}. To account for corrections beyond the leading order, higher-order pion interactions are introduced through a systematic expansion using power counting, resulting in chiral perturbation theory~\cite{Weinberg:1978kz,Gasser:1983yg,Gasser:1984gg}. The inclusion of baryons can be achieved by adding interaction terms that respect the symmetries of the theory~\cite{Weinberg:1990rz,Gasser:1987rb}. Nowadays, chiral effective theories anchored on chiral symmetry breaking have become the most popular frameworks for studying hadrons---both mesons and baryons---dynamics~\cite{Hatsuda:1994pi,Bernard:2007zu,Machleidt:2011zz}. However, this approach views the baryon as a point particle coupling to the meson field at the Lagrangian level, which falls short in addressing the detailed internal structure of baryons. This limitation has prompted the development of various theoretical models to describe baryons more comprehensively~\cite{Skyrme:1961vq,Skyrme:1962vh,Zahed:1986qz,Ma:2016npf,Capstick:1986ter,Koniuk:1979vy,Capstick:2000qj}.

Prior to the formulation of QCD, Skyrme discovered topological soliton solutions composed of non-linearly interacting pion fields, referred to as skyrmions, and proposed them as candidates for baryon models~\cite{Skyrme:1961vq,Skyrme:1962vh,Zahed:1986qz,Ma:2016npf}. This idea gained further validation in the large $N_c$ limit of QCD, where the theory simplifies and becomes dominated by planar diagrams~\cite{tHooft:1973alw}. In this limit, baryons can be effectively described as solitons of interacting meson fields because their physical properties exhibit the same $N_c$ scalings~\cite{Witten:1979kh,Adkins:1983ya,Witten:1983tw}. In $(3+1)$ dimensions, skyrmions are supported by a non-trivial homotopy group, $\pi_3(SU(N_f)) = \mathbb{Z}$ for $N_f \geq 2$, with the integer $\mathbb{Z}$ corresponding to the winding number. The winding number, conserved by virtue of topology, can be interpreted as the baryon number in the context of QCD. Nevertheless, for the case of a single quark flavor, the homotopy group becomes trivial, $\pi_3(U(1)) = 0$, implying the absence of topologically stable soliton solutions that can be associated with baryons. This limitation has been recognized as a fundamental shortcoming of the skyrmion approach to baryon physics. However, recently, the construction of the one-flavor baryon in the $(2+1)$-dimensional $\eta'$ domain wall has been suggested, which also inspires reconsideration of the chiral bag in $(3+1)$ dimensions.

In the literature, most research has focused on the dynamics of chiral mesons, particularly pions, within the chiral bag model~\cite{Rho:1983bh,Nielsen:1991bk,Rho:1993gr,Nadkarni:1985dn,Nadkarni:1984eg}. Less attention has been given to the pseudoscalar isosinglet meson $\eta'$, which acquires mass from the $U(1)_A$ anomaly~\cite{Witten:1979vv,Veneziano:1979ec,DiVecchia:1980yfw,Kawarabayashi:1980dp,Kawarabayashi:1980uh,Ohta:1981ai}. In the large $N_c$ limit, $U(1)_A$ symmetry is restored, and the $\eta'$ becomes a massless Nambu--Goldstone boson in the chiral limit. Consequently, the $\eta'$ should not be neglected when considering hadron physics in the large $N_c$ scenario and the topology of low-energy QCD at high density~\cite{Ma:2020nih}, especially given recent developments in the physics of $\eta'$ domain walls.  This prompts a reexamination of the role of the $\eta'$ in the chiral bag model.

Recently, the scope of 't Hooft anomaly matching has been extended to encompass discrete symmetries and higher-form symmetries~\cite{Gaiotto:2014kfa,Kapustin:2014lwa,Kapustin:2014gua}, providing novel insights into the properties of the $\eta'$ domain wall.  It has been shown that the $\eta^\prime$ domain wall hosts a topological $SU(N_c)_{N_f}$ Chern--Simons theory~\cite{Gaiotto:2014kfa,Gaiotto:2017yup,Gaiotto:2017tne}, which is conjectured, via level-rank duality, to be equivalent to a $U(N_f)_{-N_c}$ Chern--Simons theory~\cite{Hsin:2016blu}.  In the single-flavor scenario, baryonic states can be realized on the $\eta^\prime$ domain wall in a manner analogous to quantum Hall droplets. These droplets exhibit a spin of $N_c/2$ and support chiral edge modes, encapsulating the correct baryon number~\cite{Komargodski:2018odf}. Furthermore, the interpretation of baryons as quantum Hall droplets can be extended to view them as chiral bags within a $(2+1)$-dimensional strip, using the Cheshire Cat principle (CCP)~\cite{Ma:2019xtx}.  This provides a direct mapping between the microscopic QCD degrees of freedom, i.e., quarks and gluons, and the macroscopic degrees of freedom, i.e., hadrons, in terms of topological objects~\cite{Ma:2019ery}.

To provide a more concrete framework, it has been demonstrated that a Chern--Simons--Higgs theory emerges on the $\eta^\prime$ domain wall~\cite{Lin:2023qya}. This theory supports vortex solutions, with each vortex carrying a unit of topological charge and naturally exhibiting a spin of $N_c/2$. The $N_c$ scaling of these vortices aligns with known baryon properties, suggesting a perspective in which baryons can be interpreted as vortices.  Through particle-vortex duality, the Zhang--Hansson--Kivelson framework~\cite{Zhang:1988wy} implies that quarks carry a fractional topological charge of $1/N_c$ and follow fractional statistics.  

On the other hand, the chiral bag model has provided a successful framework for constructing baryons in the case of multiple flavors, $N_f \geq 2$~\cite{Chodos:1974je,Thomas:1982kv,Rho:1983bh,Chodos:1975ix,Mulders:1984df,Goldman:1980ww,Nadkarni:1985dm,Nadkarni:1984eg,Perry:1986sz,Damgaard:1992cy,Nadkarni:1985dn}. However, attempts to extend the chiral bag model to include an $\eta'$ boundary condition have encountered significant challenges. It has been predicted that such configurations allow color to leak from the bag, thereby breaking confinement~\cite{Nielsen:1991bk,Nielsen:1992va}.  To address this issue, a counterterm on the bag surface has been proposed, restoring gauge invariance through the $SU(N_c)_{N_f}$ Chern--Simons theory observed on the $\eta'$ domain wall~\cite{Lin:2025lzf}. This development has sparked further investigation into the interplay between the internal dynamics of the bag and the induced surface physics in chiral bag models.

In recent years, significant progress has been made in constructing baryons involving the $\eta^\prime$ meson field, and this article reviews these developments. The outline of this article is as follows: In Section \ref{sec2}, we review the fundamental aspects of the Skyrme model and chiral bag model, especially regarding topological baryon number and the color charge flow-out problem.  In Section \ref{sec3}, we examine the qualitative and semi-quantitative construction of one-flavor baryons on the $\eta'$ domain wall, where the baryon, as a quantum Hall droplet, provides insights into quark-hadron continuity. In Section \ref{sec4}, we propose the concrete theory on the $\eta'$ domain wall, where vortex solutions can be understood as baryon or multi-baryon structures based on the large $N_c$ scaling. The relevance of domain-wall skyrmions is also discussed. In Section \ref{sec5}, we explore the hypothesis that confinement is driven by monopoles confined within the bag and derive the effective theory for the bag surface. The role of vector meson fields on the bag surface is emphasized, as they are essential for ensuring the correct baryon number in the chiral bag framework. Finally, we conclude with a summary and brief discussion.

\section{Baryon Construction for \boldmath{$N_f \ge 2$}}\label{sec2}

For the case of flavor number $N_f \geq 2$, baryons have been successfully constructed using both the Skyrme model~\cite{Skyrme:1961vq,Skyrme:1962vh,Zahed:1986qz,Ma:2016npf} and the chiral bag model~\cite{Chodos:1974je,Thomas:1982kv,Rho:1983bh,Chodos:1975ix,Mulders:1984df,Goldman:1980ww,Nadkarni:1985dm,Nadkarni:1984eg,Perry:1986sz,Damgaard:1992cy,Nadkarni:1985dn}. In this section, we review these two approaches separately and explore how they lead to the construction of baryons in the one-flavor case.

\subsection{Skyrmion}
We consider QCD with $N_f \geq 2$ quark flavors. The QCD Lagrangian possesses an axial symmetry, $U(1)_A$, which is broken at the quantum level due to the axial anomaly. Consequently, the theory retains a global continuous symmetry $SU(N_f)_L \times SU(N_f)_R \times U(1)_B$. For $N_f$ within the confining range (below the conformal window), QCD exhibits confinement at low energies, and the global symmetries are spontaneously broken by the formation of a chiral condensate
\begin{equation}
	SU(N_f)_L \times SU(N_f)_R \times U(1)_B \rightarrow SU(N_f)_V \times U(1)_B.
\end{equation}
Here, $SU(N_f)_V$ is the diagonal subgroup of $SU(N_f)_L \times SU(N_f)_R$, which leaves the chiral condensate invariant. This spontaneous breaking of chiral symmetry results in massless Nambu--Goldstone bosons, identified as pions. The low-energy dynamics are captured by a non-linear sigma model, with the leading-order effective Lagrangian parametrized by $U(x) \in SU(N_f)$
\begin{equation}
	\mathcal{L} = \frac{f_\pi^2}{4} \operatorname{tr} \left( \partial_\mu U^\dagger \partial^\mu U \right) + \cdots,\quad U(x)=\exp\left\{\frac{2\mathrm{i}}{f_{\pi}}\pi^{a}(x)\lambda^{a}\right\},
\end{equation}
where higher-derivative terms, represented by $\cdots$, may be included if necessary. Notably, for $N_f \geq 3$, the Wess--Zumino term is crucial for anomaly matching~\cite{Witten:1983tw,Wess:1971yu}. It is intriguing that this effective theory, though only incorporating the massless Nambu–Goldstone (NG) modes, still allows for baryonic soliton solutions, known as skyrmions after stabilization.

To stabilize the energy of these soliton solutions, Skyrme introduced a higher-derivative term beyond leading order, forming the celebrated Skyrme model
\begin{equation}
	\mathcal{L}_{\mathrm{Sky}} = \frac{f_\pi^2}{4} \operatorname{tr} \left( \partial^\mu U^\dagger \partial_\mu U \right) + \frac{1}{32e^2} \operatorname{tr} \left( [ U^\dagger \partial^\mu U, U^\dagger \partial^\nu U ] [ U^\dagger \partial_\mu U, U^\dagger \partial_\nu U ] \right).
\end{equation}
For any soliton solution of the Skyrme model, static field configurations map spatial $\mathbb{R}^3$ to the group manifold $SU(N_f)$. To ensure finite energy, the field must reach its vacuum at spatial infinity, $\lim_{r \to \infty} U(x) = 1$, effectively compactifying $\mathbb{R}^3$ to $S^3$. Thus, soliton solutions are classified by
\begin{equation}
	\pi_3(SU(N_f)) = \mathbb{Z}, \quad \forall N_f \geq 2,
\end{equation}
and the associated topological current is the skyrmion current
\begin{equation}
	S^\mu = \frac{1}{24 \pi^2} \epsilon^{\mu \nu \rho \sigma} \operatorname{tr} \left( U^\dagger \partial_\nu U U^\dagger \partial_\rho U U^\dagger \partial_\sigma U \right),
	\label{Scurrent}
\end{equation}
which is identically conserved, $\partial_\mu S^\mu = 0$. The topological charge, given by $S = \int \mathrm{d}^3 x \, S^0$, provides a conserved quantum number. In the simplest case of $N_f = 2$, the field $U(x)$ wraps around the target space $SU(2) \cong S^3$ as one traverses spatial $\mathbb{R}^3$. This is achieved by the hedgehog ansatz~\cite{Adkins:1983ya}
\begin{equation}
	U_{\mathrm{Sky}}(\mathbf{x}) = \exp \left( i f(r) \, \boldsymbol{\sigma} \cdot \hat{\mathbf{x}} \right) = \cos f(r) + i \, \boldsymbol{\sigma} \cdot \hat{\mathbf{x}} \, \sin f(r).
\end{equation}
To satisfy the boundary condition $U(\infty) = 1$, one can choose $f(\infty) = 0$. To ensure that $U(x)$ has a well-defined limit at the origin, we also require $f(0) = n \pi$, where $n \in \mathbb{Z}$. It can then be shown that this configuration has a topological charge $S = n$.

One compelling argument for identifying skyrmions as low-energy descriptions of baryons lies in their $N_c$ scaling behavior, which matches that of baryons. Additionally, 't Hooft anomaly matching conditions further relate the skyrmion current to the baryon current, providing a strong indication that skyrmions capture essential aspects of baryonic physics~\cite{Witten:1983tw,Karasik:2022tmd}. To capture the topological characteristics of a skyrmion, one can analyze the static pion field configuration $U(x)$ as a background field. Introducing a coupling of $N_c$ quarks to this skyrmion background, we write
\begin{equation}
	\mathcal{L}_{\mathrm{Sky-quark}} = \bar{\psi} \left( \mathrm{i} \slashed{\partial} - \mu \exp\left\{ \frac{2\mathrm{i}}{f_{\pi}} \gamma^5 \pi^a(x) \lambda^a \right\} \right) \psi.
\end{equation}
Here, the specific value of the coupling constant $\mu$ is not essential for our analysis. By quantizing the system in the presence of the $U(x)$ background field, one can compute the baryon number associated with the skyrmion background. This calculation yields a baryon number that exactly equates to the topological charge $S$~\cite{Niemi:1984vz}, providing further evidence that QCD baryons correspond to skyrmions in the low-energy effective theory.

Using the skyrmion approach, not only single baryons but also nuclei---multiskyrmion state---and nuclear matter can be accessed using a unified framework~\cite{Battye:2006na,Park:2009bb,Brown:2010api,Ma:2016gdd,Naya:2018kyi,Manton:2022fcb,Wang:2023qxq}. A novel phenomenon that has not been found in other approaches than the skyrmion one is the topology change where the matter made of one-winding number objects transit to that made of half-winding number objects~\cite{Kugler:1988mu,Kugler:1989uc,Goldhaber:1987pb,Lee:2003eg}. In terms of CCP, this topology change represents the hadron–quark continuity in QCD in hadronic matter~\cite{Ma:2019xtx,Ma:2019ery}. It has been found that this topology change is essential for the equation of state for compressed baryonic matter relevant for massive compact stars~\cite{Lee:2010sw,Holt:2007ih,Paeng:2015noa,Paeng:2017qvp,Ma:2018jze}.

\subsection{Chiral Bag}

\label{subchiral}

The skyrmion description of baryons emphasizes global properties but does not capture their quark composition. To address this, the chiral bag model of baryons was proposed as an extension of the MIT bag model~\cite{Chodos:1974je,Thomas:1982kv,Rho:1983bh,Chodos:1975ix,Mulders:1984df,Goldman:1980ww}, incorporating spontaneous chiral symmetry breaking. In this model, the MIT bag is surrounded by a cloud of chiral mesons, such as pions. The bag represents a static three-dimensional region $\Omega$, bounded by a smooth surface $\Sigma$. Here, $\Sigma$ is not required to be connected, and $\Omega$ need not be simply connected. Inside the bag, quarks are free and satisfy the Dirac equation $i \gamma^\mu \partial_\mu \psi = 0$. To confine the quarks, $\psi$ obeys a boundary condition on $\Sigma$~\cite{Hosaka:1996ee}
\begin{equation}
	-i \bm{\gamma} \cdot \bm{n} \psi = \exp(i \theta_\Sigma \bm{\sigma} \cdot \bm{n} \gamma_5) \psi ,
\end{equation}
where we consider two-flavor quarks, with $\bm{\sigma}$ as the Pauli matrices, and $\bm{n}$ the unit exterior normal to $\Sigma$. The bag surface is parametrized by the chiral angle $\theta_\Sigma$. In the chiral bag model, the baryon number is defined as
\begin{equation}
	N_B = -\frac{1}{2} \lim_{t \to +0} \sum_n \epsilon(E_n) \exp(-t |E_n|),
\end{equation}
where the sum extends over all positive and negative energy single-particle eigenstates. This regulated expression corresponds to $\int d^3x \frac{1}{2} [\psi^\dagger(x), \psi(x)]$. Using the Dirac equation and boundary conditions, it can be shown that~\cite{Goldstone:1981kk,Jaroszewicz:1983wd}
\begin{equation}
	\frac{dN_B}{d\theta_\Sigma} = \frac{1}{2} \lim_{t \to +0} t \sum_n \frac{dE_n}{d\theta_\Sigma} \exp(-t |E_n|) = \frac{1}{\pi} \chi \sin^2 \theta_\Sigma,  \label{dN}
\end{equation}
where $\chi$ is the Euler characteristic of $\Sigma$, given by twice the number of pieces of $\Sigma$ minus twice the number of handles. 

If chiral symmetry is unbroken within the bag, the chiral angle must take a vacuum value, $\theta_\mathrm{in}$. Moving from the bag's interior to its exterior across $\Sigma$ causes a discontinuous change in the chiral angle from $\theta_\mathrm{in}$ to $\theta_\Sigma$. For a spherical bag surface $\Sigma$, $\chi=2$, we can integrate ${dN}/{d\theta_\Sigma}$ in Equation~(\ref{dN}) to determine the baryon number inside the two-flavor chiral bag
\begin{equation}
	N_\mathrm{in} = \frac{1}{\pi} \left\{ \theta_\Sigma - \theta_\mathrm{in} - \frac{1}{2} \left[ \sin 2 \theta_\Sigma - \sin 2 \theta_\mathrm{in} \right] \right\}.
\end{equation}

Outside the chiral bag, the baryon number is carried by the meson cloud, which is described by the Skyrme model. Adopting the hedgehog ansatz and setting the chiral angle at infinity to $\theta_\infty$, we insert $U = \exp(i \theta \bm{\tau} \cdot \bm{n}_\beta)$ into the skyrmion current Equation~(\ref{Scurrent}), and integrate from the bag surface to infinity to obtain the baryon number outside the chiral bag
\begin{equation}
N_\mathrm{out} = \frac{1}{\pi} \left\{ \theta_\infty - \theta_\Sigma - \frac{1}{2} \left[ \sin 2 \theta_\infty - \sin 2 \theta_\Sigma \right] \right\}.
\end{equation}
The total baryon number of the chiral bag is thus
\begin{equation}
N = N_\mathrm{in} + N_\mathrm{out} = \frac{1}{\pi} \left\{ \theta_\infty - \theta_\mathrm{in} - \frac{1}{2} \left[ \sin 2 \theta_\infty - \sin 2 \theta_\mathrm{in} \right] \right\},
\end{equation}
which precisely matches the result from the topological soliton bag model for baryons~\cite{Rho:1983bh}.

There are also gluons present within the bag. To confine color charge within the bag, gluons must satisfy the following boundary conditions on the bag surface~\cite{Chodos:1974je}
\begin{equation}
\boldsymbol{n} \cdot \boldsymbol{E}^a_G = 0, \quad \boldsymbol{n} \times \boldsymbol{B}^a_G = 0, \label{BD1}
\end{equation}
where $\bm{n}$ is the outward normal to the bag surface $\Sigma$, $\boldsymbol{E}^a_G = F_G^{a,i0}\boldsymbol{e}_i$ represents the color electric field tangent to the surface, and $\boldsymbol{B}^a_G = -\frac{1}{2} \epsilon^{ijk} F_{G,jk} \boldsymbol{e}_i$ represents the color magnetic field orthogonal to the surface, $\boldsymbol{e}_i$ is the unit vector in $i= 1,2,3$ direction. Here, $a, b, c$ are color indices. Under these boundary conditions, quarks are restricted to move along the bag surface and cannot escape at the classical level. However, at the quantum level, this scenario changes.

So far, we have discussed the boundary conditions involving only pion fields in the bag model. However, the pseudoscalar isosinglet meson field $\eta^\prime$ is also important for a more comprehensive picture. The coupling of $\eta^\prime$ to quarks on the bag boundary is formulated as follows~\cite{Nadkarni:1985dm,Hosaka:1996ee}:
\begin{equation}
\left[\mathrm{i} \gamma \cdot \bm{n} + \mathrm{e}^{\mathrm{i} \gamma_5 {\eta^\prime_\Sigma}/{f_{\eta^\prime}}}\right] \psi = 0, \label{Boun}
\end{equation}
where $f_{\eta^\prime}$ is the decay constant of $\eta^\prime$. For convenience, we redefine the $\eta^\prime$ field by $\eta^\prime \rightarrow \eta^\prime f_{\eta^\prime}$ to absorb the decay constant in the following, so that $\eta^\prime$ becomes a dimensionless field. It has been observed, however, that allowing for a pseudoscalar singlet coupling at the bag surface can result in color charge leakage~\cite{Nielsen:1991bk}.

Within the bag, the color charge $Q^a_G$ is defined through the QCD current as
\begin{equation}
Q^a_G = \int_{\Omega} \mathrm{d}^3x \, j_0^a = \int_{\Omega} \mathrm{d}^3x \left( g\psi^\dagger \frac{1}{2} \lambda^a \psi + g f^{abc} G_i^b E^{ci} \right) \simeq \oint_{\Sigma} \mathrm{d}S \, \boldsymbol{E}^a_G \cdot \bm{n},
\end{equation}
where the final expression follows from Gauss' law. Here, $\lambda^a$ represents the Gell-Mann matrices, and $f^{abc}$ are the structure constants of the $SU(3)$ color algebra. Although the QCD action is locally gauge-invariant within the bag, color charge $Q^a$ is not conserved at the quantum level. In the case of one-flavor quarks and quasi-Abelian symmetry, color charge leakage can occur when the $\eta^\prime$ field varies with time~\cite{Nielsen:1991bk}
\begin{equation}
\frac{\mathrm{d}Q^a_G}{\mathrm{d}\eta^\prime_\Sigma} = \frac{g^2}{8\pi^2} \oint_{\Sigma} \mathrm{d}S \, \boldsymbol{B}^a_G\cdot \bm{n} \simeq \frac{\mathrm{d}}{\mathrm{d}\eta^\prime_\Sigma} \oint_{\Sigma} \mathrm{d}S \, \boldsymbol{E}^a_G \cdot \bm{n}. \label{dQ}
\end{equation}
This relationship is argued to apply to non-Abelian color magnetic fields as well. Such color charge leakage is problematic as it breaks both confinement and gauge invariance. To counteract this, a gauge-dependent counterterm has been proposed~\cite{Nielsen:1991bk}
\begin{equation}
S_{\mathrm{CT}} = \frac{g^2 N_f}{16\pi^2} \oint_{\Sigma} \mathrm{d}\beta \, \eta^\prime_\Sigma \, n_{\mu} K_5^\mu, \quad K_5^\mu = \epsilon^{\mu\nu\alpha\beta} \left( G_\nu^a G_{\alpha\beta}^a - \frac{2}{3} f^{abc} g G_\nu^a G_\alpha^b G_\beta^c \right). \label{CT}
\end{equation}
If we choose the vacuum value of $\eta^\prime$ inside the bag as $\eta^\prime_{\mathrm{in}}$, which shifts sharply to $\eta^\prime_{\Sigma}$ on the bag surface, we can express the counterterm more explicitly as
\begin{equation}
\frac{\eta^\prime_{\Sigma} - \eta^\prime_{\mathrm{in}}}{2\pi} \times \int_{\Sigma} \frac{N_f}{4\pi} \left( G \, \mathrm{d}G - \mathrm{i} \frac{2}{3} G^3 \right),
\label{CTCS}
\end{equation}
where differential form notation is used, and the gauge coupling is absorbed into the gauge field. This expression reveals a Chern--Simons theory arising on the bag surface. This result shows a striking similarity to recent studies on $\eta^\prime$ domain walls, where a similar topological field theory is predicted~\cite{Gaiotto:2017yup,Gaiotto:2017tne}. This leads us to suggest that the Chern--Simons theory on the chiral bag surface may be analogous to that on the $\eta^\prime$ domain wall.

The construction of baryons discussed above holds primarily for multi-flavor cases ($N_f \geq 2$). How to understand one-flavor baryons from a topology point of view remained a puzzle for many years. A recent proposal to construct one-flavor baryon using $\eta^\prime$ domain walls in $(2+1)$ dimensions~\cite{Komargodski:2018odf} put forward this direction and deepens our understanding of the chiral bag model~\cite{Ma:2019xtx,Karasik:2020pwu,Rho:2021aad,Karasik:2020zyo,Lin:2023qya}.

% Modifications:
% 1. "beyond the leading order" → "beyond leading order" for improved flow.
% 2. Removed unnecessary comma before citation.
% 3. "is formulated" instead of "can be formulated" for active voice.
% 4. Added space before integral symbols for consistency.
% 5. Removed "a" from "a striking similarity" for smoother sentence flow.

\section{Baryons as Quantum Hall Droplets and Quark-Hadron Duality}\label{sec3}

For the multi-flavor case $N_f \geq 2$, we have reviewed the construction of baryons through two approaches: the Skyrmion model and the chiral bag model. Both are based on the Nambu--Goldstone bosons arising from chiral symmetry breaking and the non-trivial homotopy group $\pi_3(SU(N_f)) = \mathbb{Z}$, valid for all $N_f \geq 2$. However, this approach does not work in the one-flavor case. The $\eta^\prime$ meson acquires a significant mass due to the axial anomaly, so there is no Nambu--Goldstone boson present~\cite{tHooft:1976rip,tHooft:1986ooh}. Even in the large $N_c$ limit, where the $U_A(1)$ symmetry is restored and the $\eta^\prime$ meson becomes massless, the homotopy group $\pi_3(U(N_f=1)) = 0$ is trivial, making it impossible to construct a one-flavor baryon in $(3 + 1)$ dimensional spacetime. However, if we consider the one-flavor baryon in $(2 + 1)$ dimensions, a construction is possible. Such a construction takes place on an $\eta^\prime$ domain wall, exhibiting several unique properties~\cite{Komargodski:2018odf,Ma:2019xtx,Lin:2023qya,Karasik:2020zyo,Karasik:2020pwu}.

As mentioned, in the large $N_c$ limit, the $U_A(1)$ symmetry becomes exact, and its breaking leads to the emergence of a Nambu--Goldstone boson, the $\eta^\prime$. The $\eta^\prime$ is a periodic scalar, satisfying $\eta^\prime \sim \eta^\prime + 2\pi$. The effective Lagrangian, which includes the leading $1/N_c$ correction, is given by~\cite{Witten:1979vv,Veneziano:1979ec}
\begin{equation}
	\mathcal{L}^{\mathrm{eff}}_{\eta^\prime} = \frac{N_f f_\pi^2}{8} \, \mathrm{d}\eta^\prime \wedge \star \mathrm{d}\eta^\prime + \frac{f_\pi^2}{8 N_f} m_{\eta^\prime}^2 \min_{n \in \mathbb{Z}} \left( N_f \eta^\prime + \theta - 2\pi n \right)^2, \label{Leta}
\end{equation}
where the potential term is locally quadratic but has a cusp whenever $\eta^\prime = \pi \, \mathrm{mod} \, 2\pi$. Physically, this cusp implies that when $\eta^\prime$ crosses $\pi$, heavy fields transition between vacua. The domain wall between $\eta^\prime = 0$ and $\eta^\prime = \pi$ is a metastable configuration~\cite{Komargodski:2018odf}. This domain wall supports a topological field theory on its world volume $\mathcal{M}_3$, which has been identified as an $SU(N_c)_{N_f}$ Chern--Simons (CS) theory
\begin{equation}
	\frac{i}{4\pi}\int_{\mathcal{M}_3} N_f \mathrm{Tr}\left(\widetilde{a}\wedge d\widetilde{a}+\frac{2}{3}\widetilde{a}^3\right),
\end{equation}
where $\tilde{a}$ is the $\mathfrak{su}(N_c)$-valued gauge field, $N_f$ and $N_c$ are referred to as the level and rank of the Chern--Simons (CS) theory, respectively.

\begin{figure}[H]
	\centering  %图片全局居中
	\subfigure[]{
		\label{Pancake}
		\includegraphics[width=0.45\textwidth]{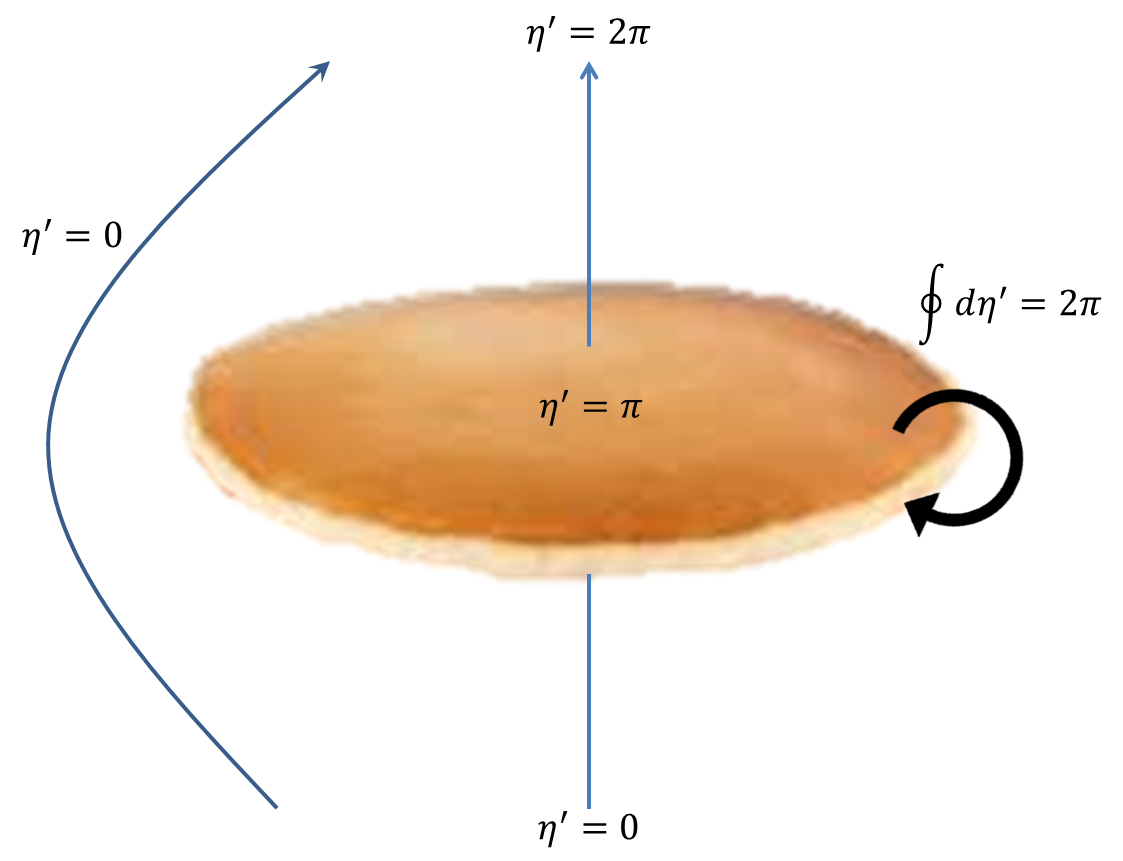}}
	\subfigure[]{
		\label{droplet}
		\includegraphics[width=0.45\textwidth]{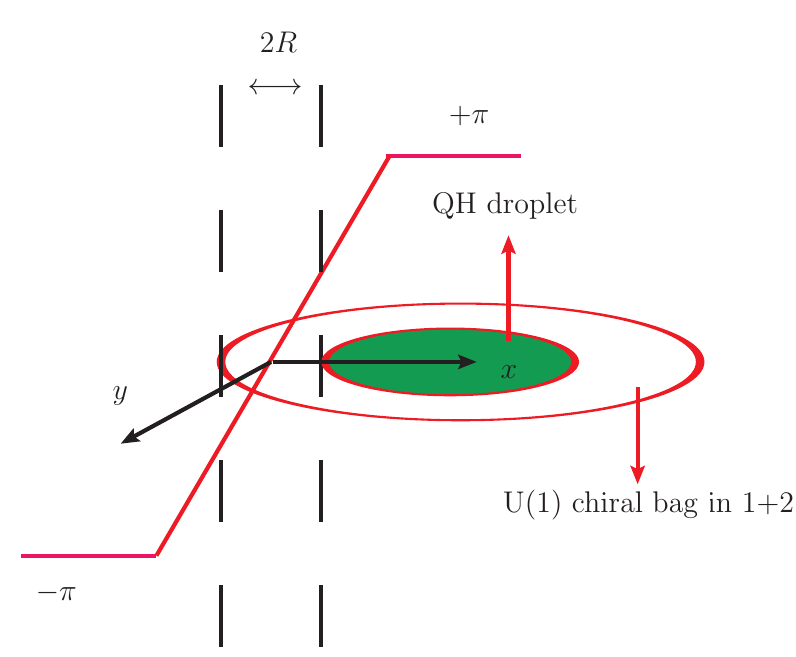}}
	\caption{Baryons as quantum Hall droplets and quark-hadron duality. (\textbf{a}) The one-flavor baryon is suggested to be a quantum Hall droplet (picture from~\cite{Karasik:2020pwu}). (\textbf{b}) The one-flavor baryon is constructed from the chiral bag model by the Cheshire Cat principle (picture from~\cite{Ma:2019xtx}).}
	\label{fig1}
\end{figure}

To understand how this theory couples to the background baryon gauge field $A$, we apply level-rank duality~\cite{Naculich:1990pa,Naculich:2007nc,Camperi:1990dk,Nakanishi:1990hj,Aharony:2015mjs}:  
\begin{equation}
	SU(N_c)_{N_f} \longleftrightarrow U(N_f)_{-N_c}.
\end{equation}
For the one-flavor case with $N_f = 1$, level-rank duality transforms the non-Abelian theory into an Abelian one, effectively gauging the flavor symmetry. This allows us to introduce a $\mathfrak{u}(1)$ gauge field $a$ and write the action on the domain wall as
\begin{equation}
	\int_{\mathcal{M}_3} \frac{N_c}{4\pi} a \wedge \mathrm{d}a + \frac{1}{2\pi} a \wedge \mathrm{d}A.
\end{equation}
Here, level-rank duality transforms the baryon symmetry in the $SU(N_c)_{1}$ description into magnetic symmetry in the $U(1)_{-N_c}$ description. This action resembles that of the fractional quantum Hall effect, where imposing a boundary on the domain wall gives rise to a quantum Hall droplet shown in Figure~\ref{fig1}a. This droplet is suggested to represent a baryon, with chiral boundary excitations, where specific boundary vertex operators correspond to states with baryon numbers. If the chiral boundary mode carries a unit baryon number, then the droplet possesses a spin of $N_c/2$, matching that of a one-flavor baryon. Additionally, the mass, size, and excitation properties of this droplet align with those of a one-flavor baryon~\cite{Komargodski:2018odf}.

The chiral bag construction further clarifies details on baryons as quantum Hall droplets. As mentioned above, the boundary chiral mode is essential, carrying baryon number and inducing spin. However, the mechanism by which quarks form baryons remained unclear until it was interpreted via the Cheshire Cat principle (CCP). As shown in Figure~\ref{fig1}b, the boundary of the quantum Hall droplet can be extended to an annular-shaped chiral bag, with a width of $2R$, where the $\eta^\prime$ field forms a monodromy of $2\pi$. For a single-flavor quark species, the chiral bag model on the annulus is described by~\cite{Zahed:1984jmf}
\begin{equation}
	\begin{aligned}
		(i\partial_t + i\sigma_2 \partial_x - i\sigma_3 \partial_y) q(t, x, y) &= 0, \quad |x| < R, \\
		(e^{-i\sigma_2 \theta(t, x)} - \sigma_3 \epsilon(x)) q(t, x, y) &= 0, \quad |x| = R,
	\end{aligned}
\end{equation}
where $\epsilon(x) = x/|x|$ is the outward normal to the bag, and $(\gamma^{0}, \gamma^{1}, \gamma^{2}) = (\sigma_{1}, i\sigma_{3}, i\sigma_{2})$. The $\eta^\prime$ field acts at the boundary through the chiral angle $\theta = \eta^{\prime}/f_\eta$, which can be time-dependent but is independent of $y$. Here, $f_\eta$ is the decay constant of $\eta^\prime$. Solving these equations yields a twisted spectrum
\begin{equation}
	E_n = \frac{(2n+1)\pi}{4R} + \frac{\Delta \theta}{4R}, \quad \Delta \theta = [\theta(+R) - \theta(-R)],
\end{equation}
where $\Delta \theta$ represents the jump in the $\eta^\prime$ field across the chiral bag. At the ``magic angle'' $\Delta \theta = \pi$, the energy level $E_{-1}$ crosses zero, requiring a redefinition of the vacuum. This redefinition implies that the baryon charge $Q$ fractionalizes into three parts, each carried by the Hall droplet, the chiral bag, and the exterior of the bag~\cite{Niemi:1984vz,Goldstone:1981kk,Zahed:1984jmf}
\begin{equation}
	Q = \frac{e}{2\pi} \left\{ \left[\theta(-R) - (-\pi)\right] + \Delta \theta + \left[\pi - \theta(+R)\right] \right\} = e,
\end{equation}
where $e$ is the total baryon charge. According to the CCP, physical properties should not depend on the bag size. As the bag radius decreases, the bag reduces to a vortex line, resembling the ``smile'' of the Cheshire Cat, with gapless quarks circulating in the same direction. The disk enclosed by the ``smile'' is described by a topological field theory due to the Callan--Harvey anomaly outflow~\cite{Callan:1984sa}
\begin{equation}
	\int_{\mathcal{M}_3} \frac{N_c}{4\pi} a \wedge \mathrm{d}a,
\end{equation}
which exactly corresponds to the $U(1)_{-N_c}$ Chern--Simons field introduced earlier. Thus, inside the vortex line, quarks carry baryon charge and are in the fundamental representation of the gauge group $SU(N_c)_1$. Outside, the gauge field is $U(1)_{-N_c}$, with its flux serving as a measure of baryon charge. These two Chern--Simons theories are connected in the vortex line via level-rank duality. This configuration allows the vortex line to expand or contract arbitrarily without affecting the physics, implying quark-hadron continuity.

% Modifications:
% 1. Corrected usage of commas and conjunctions: "However, this approach does not work in the one-flavor case."
% 2. Rephrased the sentence for clarity: "Such a construction takes place on an $\eta^\prime$ domain wall, exhibiting several unique properties."
% 3. Corrected "a Nambu-Goldstone" to "an Nambu-Goldstone" for grammatical correctness.
% 4. Adjusted sentence structure to improve clarity and readability: "Physically, this cusp implies that when $\eta^\prime$ crosses $\pi$, heavy fields transition between vacua."
% 5. Reworded for fluency: "level-rank duality transforms the baryon symmetry in the $SU(N_c)_{1}$ description into magnetic symmetry in the $U(1)_{-N_c}$ description."

\section{Baryons as Vortices}\label{sec4}

The one-flavor baryons, interpreted as quantum Hall droplets under quark-hadron duality on the $\eta^\prime$ domain wall, primarily involve qualitative insights, leaving a complete theoretical formulation as an open challenge. Although the level-rank duality in Chern--Simons theory provides a foundational aspect, it is insufficient. For multiple flavors, the Chern--Simons (CS) field $SU(N_c)_{N_f}$ present on the domain wall is expected to correspond to the gluon field. Advancing towards a comprehensive theory requires incorporating strongly interacting quarks or fermions within the framework.

In this context, the dual theory becomes more complex with the introduction of additional fields. The proposed duality between the two formulations can be written as~\cite{Gaiotto:2017tne,Hsin:2016blu}
\begin{equation}
	SU(N_c)_{-N_f} + N_f \; \mathrm{fermions} \quad \longleftrightarrow\quad  U(N_f)_{N_c} + N_f \; \mathrm{scalars}.
	\label{dualityNf}
\end{equation}
On the right-hand side, the dual theory incorporates $N_f$ scalar fields, representing the bosonized form of the fermions or quarks from the left-hand side. We should admit that we cannot explicitly identify the physical content of the $N_f$ scalars but as auxiliary fields in the $CP^{N_f-1}$ model. Duality~\eqref{dualityNf} establishes a link between the $SU(N_c)$ color gauge symmetry with $N_f$ fermions in the fundamental representation and the $U(N_f)$ baryon gauge symmetry with $N_f$ scalar fields in the fundamental representation. In this framework, the global flavor symmetry $U(N_f)$ becomes gauged through level-rank duality. Such a setup provides a pathway for developing a concrete theory on the domain wall, where baryons appear as vortices, analogous to skyrmions viewed as topological solitons.

\subsection{Baryons as Vortices for $N_f=1$ and Particle-Vortex Duality}

To simplify the discussion, we begin with the one-flavor case, i.e., $N_f = 1$, and later generalize to multi-flavor. For $N_f = 1$, the duality on the $\eta'$ domain wall described in Equation~(\ref{dualityNf}) simplifies to
\begin{equation}
	SU(N_c)_{-1} + \mathrm{one\;fermion}\;\psi \quad\longleftrightarrow\quad  U(1)_{N_c} + \mathrm{one\;scalar}\;\phi.
	\label{duality1}
\end{equation}
On the right-hand side, the effective field theory consists of a $U(1)_{N_c}$ Chern--Simons (CS) gauge field $a_\mu$ and a complex dual scalar field $\phi$. Assuming gauge invariance and minimal coupling, the effective Lagrangian in $(2+1)$-dimension can be conjectured as
\begin{equation}
	\mathcal{L}_A[\phi, a] = \int d^3x \, |\partial_\mu\phi - \mathrm{i}a_\mu\phi|^2 + \frac{N_c}{4\pi} \epsilon^{\mu\nu\rho} a_\mu \partial_\nu a_\rho - V(\phi^*\phi),
	\label{LA}
\end{equation}
where $\phi$ represents the bosonized form of the fermion $\psi$, i.e., $\phi^*\phi \sim \psi^\dagger\psi$. The gauge field $a_\mu$ is an emergent $U(1)$ field linked to baryon number conservation before gauging. This Lagrangian~\eqref{LA} is consistent with the quantum Hall description of baryons, where the fermion $\psi$ (quark) replaces the electron in fractional quantum Hall systems at a $1/N_c$ filling fraction. In this context, $\phi^*\phi$ corresponds to the quark density, and $a_\mu$ propagates the quark number. Since quarks carry color charge, the quark density per color, $\phi_c^*\phi_c = \phi^*\phi / N_c$, must remain finite in the large $N_c$ limit.

The potential $V(\phi^*\phi)$ obtains a non-zero vacuum expectation value $\langle \phi^*\phi \rangle = N_c v^2$. It can be formally expressed as
\begin{equation}
	V(\phi^*\phi) = N_c \sum_{I=1}^\infty c_I \left(\frac{\phi^*\phi}{N_c} - v^2\right)^I, \;\;\; v > 0,
\end{equation}
where the coefficients $c_I$ are constrained to ensure that $\phi_c^*\phi_c$ has a non-zero vacuum expectation value $v^2$, independent of $N_c$.

The Lagrangian $\mathcal{L}_A$ scales with $N_c$, representing the leading contribution in the $N_c$ expansion. While additional terms of lower order in $N_c$ can be included, their effects are negligible in the large $N_c$ limit.

For the theory described by $\mathcal{L}_A$, (2+1)-dimensional topologically non-trivial finite-energy vortex solutions exist, satisfying the equations of motion
\begin{equation}
	\begin{gathered}
		(\partial_\mu - \mathrm{i}a_\mu)(\partial^\mu - \mathrm{i}a^\mu)\phi + \frac{\partial V}{\partial \phi^*} = 0, \\
		(\partial_\mu + \mathrm{i}a_\mu)(\partial^\mu + \mathrm{i}a^\mu)\phi^* + \frac{\partial V}{\partial \phi} = 0, \\
		\mathrm{i}(\phi^* \partial^\mu \phi - \phi \partial^\mu \phi^*) + 2 a^\mu \phi \phi^* + \frac{N_c}{2\pi} \epsilon^{\mu\nu\rho} \partial_\nu a_\rho = 0.
	\end{gathered}
\end{equation}
In this review, we focus on the topological properties of the vortices without solving these complex equations of motion explicitly. Additionally, we only consider the global quantities such as spin and topological charge of vortices in the large $N_c$ limit, therefore, the precise form of $V(\phi^* \phi)$ which relates to local properties for example the vertex mass, radius, etc, is not essential. For the present qualitative discussion, we only require that the potential is of the Higgs type. Considering a single vortex located at the origin, we adopt the following ansatz in polar coordinates~\cite{Jatkar:1989sc}
\begin{equation}
	\phi(\mathbf{r}) = \mathrm{e}^{\mathrm{i}n\theta} h(r), \;\; a_0(\mathbf{r}) = a_0(r), \;\; \mathbf{a}(\mathbf{r}) = \frac{a(r)}{r}(\sin\theta, -\cos\theta),
	\label{polar}
\end{equation}
with boundary conditions for finite energy
\begin{equation}
	h(\infty) =\sqrt{N_c} v, \;\; a_0(\infty) = 0, \;\; a(\infty) = n; \quad
	h(0) = 0, \;\; a_0(0) = c, \;\; a(0) = 0,
	\label{BC}
\end{equation}
where $c$ is a non-zero constant. The winding number $n \in \mathbb{Z}$ characterizes the vortex solutions, as we will discuss.

The Chern--Simons term in the Lagrangian is topological and contributes to a topological current given by
\begin{equation}
	j^\mu = \frac{N_c}{2\pi} \epsilon^{\mu\nu\rho} \partial_\nu a_\rho = \frac{N_c}{4\pi} \epsilon^{\mu\nu\rho} f_{\nu\rho},
	\label{eq:TopCharge1f}
\end{equation}
where $f_{\mu\nu} = \partial_\mu a_\nu - \partial_\nu a_\mu$ is the field strength tensor of $a_\mu$. Using this expression, the vortex solution can be shown to carry a topological charge
\begin{equation}
	Q = \int j^0 \, dx \, dy = \frac{N_c}{2\pi} \int \epsilon^{0\nu\rho} \partial_\nu a_\rho \, dx \, dy = n N_c,
\end{equation}
which corresponds to the quantization of the vortex flux
\begin{equation}
	\Phi = \int \epsilon^{0\nu\rho} \partial_\nu a_\rho \, dx \, dy = \int \mathbf{a} \cdot d\mathbf{r} = \int \frac{n}{r} r d\theta = 2\pi n.
\end{equation}

Objects that carry both flux and charge, such as these vortex solutions, are known as anyons, which obey fractional statistics~\cite{Wilczek:1981du,Rao:1992aj}. The spin of such a vortex is given by
\begin{equation}
	s = \frac{Q \Phi}{4\pi} = n^2 \frac{N_c}{2}, \quad n \in \mathbb{Z}.
\end{equation}
For vortices with $n = \pm 1$, the spin matches that of one-flavor baryons in their ground state. Since the scalar field $\phi$ corresponds to the quark number and its coupling to $a_\mu$ is normalized to unity, the topological charge $Q$ can be interpreted as the quark number. Given that a baryon consists of $N_c$ quarks, the baryon number can naturally be defined as
\begin{equation}
	B = \frac{Q}{N_c} = n.
\end{equation}
Thus, it is reasonable to conjecture that vortices with $n = \pm 1$ represent (anti)baryons on the domain wall. These vortices exhibit properties consistent with one-flavor baryons in the large $N_c$ limit. A special case is the $n = 0$ vortex, which has a zero baryon number. Given the relation $\phi^*\phi \sim \psi^\dagger\psi$, this vortex is associated with quark condensates. Vortices with $|n| \geq 2$ can be interpreted as multi-baryon structures on the domain wall, analogous to skyrmions with higher baryon numbers in $(3+1)$ dimensions~\cite{Weigel:1986zc,Houghton:1997kg}.

In the large $N_c$ limit, the quark density per-color, $\phi_c^*\phi_c = \phi^*\phi / N_c$, must remain finite. The Lagrangian in Equation~\eqref{LA} can be reformulated as
\begin{equation}
	\mathcal{L}_A = N_c  \left[ \left|\partial_\mu \phi_c - \mathrm{i} a_\mu \phi_c\right|^2 - V_c(\phi_c^* \phi_c) + \frac{1}{4\pi} \epsilon^{\mu\nu\rho} a_\mu \partial_\nu a_\rho \right],
\end{equation}
where
\begin{equation}
	V_c(\phi_c^* \phi_c) = \frac{V(\phi^*\phi)}{N_c} = \sum_{I=1}^\infty c_I \left(\phi_c^* \phi_c - v^2\right)^I, \quad v > 0.
\end{equation}
This potential is independent of $N_c$, ensuring that $N_c$ appears only as an overall factor in the Lagrangian. Consequently, $N_c$ does not enter the equations of motion or affect the vortex solutions. Therefore, the sizes of the vortices are determined solely by color-independent parameters. Following the reasoning in Ref.~\cite{Witten:1979kh}, one can deduce that vortex radii scale as $\sim O(N_c^0)$, their energy (mass) scales as $\sim O(N_c)$, and vortex-vortex scattering amplitudes are of order $N_c$. 

These scaling properties are identical to those of baryons, reinforcing the identification of vortices as baryons on the $\eta^\prime$ domain wall.

As we have seen, baryons can be described as vortices, and it is also possible to estimate the physical quantities of mesons. In large $N_c$ QCD, meson operators are of the form $\mathcal{J}(x) = \sqrt{N_c} \bar{\psi} G^m \psi$, where $G^m$ denotes any number of gluon field strengths, derivatives, and Dirac gamma matrices, and $\psi$ is the quark field. The normalization factor $\sqrt{N_c}$ ensures that the two-point function $\langle \mathcal{J} \mathcal{J} \rangle \sim O(N_c^0)$, indicating that meson operators create meson states with amplitudes of order $N_c^0$. 

On the domain wall, within the effective theory $\mathcal{L}_A$, meson operators should correspond to $\hat{\mathcal{J}}(x) = \sqrt{N_c} \phi^* f^m \phi$, where $f^m$ involves combinations of Chern--Simons field strengths, derivatives, and other operators. We can estimate the two-point function of meson operators using the quark density relation form $\phi^*\phi \sim \psi^\dagger\psi$. The coupling between two meson operators is determined by the first term of the Higgs-type potential
\begin{equation}
	c_2 N_c \left( \frac{\phi^*\phi}{N_c} - v^2 \right)^2 \sim c_2 \left( \sqrt{N_c} \phi_c^* \phi_c \right)^2 \sim c_2 \hat{\mathcal{J}} \hat{\mathcal{J}}.
\end{equation}
This leads $\langle \hat{\mathcal{J}} \hat{\mathcal{J}} \rangle \sim c_2 \sim (N_c^0)$. Similarly, the four-point function for mesons can be estimated as
\begin{equation}
	c_4 N_c \left( \frac{\phi^*\phi}{N_c} - v^2 \right)^4 \sim \frac{c_4}{N_c} \left( \sqrt{N_c} \phi_c^* \phi_c \right)^4 \sim \frac{c_4}{N_c} \hat{\mathcal{J}} \hat{\mathcal{J}} \hat{\mathcal{J}} \hat{\mathcal{J}}.
\end{equation}
Thus, the amplitude for meson-meson scattering scales as $1/N_c$, consistent with the expected behavior of large $N_c$ QCD. Higher-order correlation functions can be analyzed similarly, showing that meson properties on the domain wall align with those in large $N_c$ QCD.

To investigate baryon--meson interactions in the large $N_c$ limit, one can insert a meson operator $\hat{\mathcal{J}}(x) = \sqrt{N_c} \phi^* f^m \phi$ into a vortex solution. When the operator is inserted far from the vortex center, the stability of the vortex is preserved. Local fluctuations in the quark density occur, and the dominant energy contribution arises from the potential term, which scales as $c_2 \sim N_c^0$. This implies that the meson-vortex (baryon) scattering amplitudes are of order $N_c^0$.

For one flavor, the gauge field $a_\mu$ in the theory $\mathcal{L}_A$ arises purely from the topological Chern--Simons term, with no self-interaction term. Consequently, there is no equivalent description of glueballs in $(2+1)$ dimensions. Additionally, the requirement $\phi_c^* \phi_c = \phi^* \phi / N_c$ ensures that the kinetic term remains finite and does not survive in the large $N_c$ limit, suppressing glueball-like structures for any number of flavors.

So far, we have successfully constructed topological baryons as vortices on the $\eta^\prime$ domain wall. However, the precise role of quarks in this picture remains unclear. This ambiguity relates to the concept of particle-vortex duality~\cite{Peskin:1977kp}. In $(2+1)$ dimensions, it is possible to describe the same physics using dual theories, where particles in one theory correspond to vortices in the other, and vice versa.

For the theory $\mathcal{L}_A$, two types of excitations emerge: quantized $\phi$, representing quarks, and vortices. According to particle-vortex duality, the dual theory, known as the relativistic Zhang--Hansson--Kivelson (ZHK) theory~\cite{Zhang:1988wy,Tong:2016kpv}, is given by
\begin{equation}
	\mathcal{L}_B[\tilde{\phi}, \tilde{a}] = \int d^3x \, \left[ |\partial_\mu \tilde{\phi} - i \tilde{a}_\mu \tilde{\phi}|^2 - \tilde{V}(\tilde{\phi}^* \tilde{\phi}) + \frac{1}{4\pi N_c} \epsilon^{\mu\nu\rho} \tilde{a}_\mu \partial_\nu \tilde{a}_\rho \right],
	\label{LB}
\end{equation}
where $\tilde{\phi}$ is a complex scalar field, $\tilde{a}_\mu$ is a $U(1)$ gauge boson, and $\tilde{V}(\tilde{\phi}^* \tilde{\phi})$ is a Higgs-type potential. By particle-vortex duality, $\tilde{\phi}$ represents baryons, while vortices in $\mathcal{L}_B$ correspond to quarks or multi-quark structures. The dual theory $\mathcal{L}_B$ also admits vortex solutions, with a topological current given by
\begin{equation}
	\tilde{j}^\mu = \frac{1}{2\pi N_c} \epsilon^{\mu\nu\rho} \partial_\nu \tilde{a}_\rho.
\end{equation}
The corresponding topological charge is
\begin{equation}
	\tilde{Q} = \int \tilde{j}^0 \, dx \, dy = \frac{1}{2\pi N_c} \int \epsilon^{0\nu\rho} \partial_\nu \tilde{a}_\rho \, dx \, dy = \frac{n}{N_c},
\end{equation}
using the same parameterizations and boundary conditions as in Equations~\eqref{polar} and~\eqref{BC}. With $\tilde{\phi}^* \tilde{\phi}$ representing baryon density and $\tilde{a}_\mu$ propagating unit baryon numbers, the baryon number is defined as the topological charge $\tilde{Q}$.

Vortices in $\mathcal{L}_B$ exhibit fractional topological charge and spin:

\begin{equation}
	\tilde{s} = \frac{\tilde{Q} \Phi}{4\pi} = \frac{n^2}{2N_c}, \quad n \in \mathbb{Z}.
\end{equation}
Basic vortex states with winding numbers $\pm 1$ correspond to (anti)quarks, with the same statistics as quarks leaked from chiral bags~\cite{Ma:2019xtx}. Higher winding numbers describe multi-quark structures. Notably, vortices with $n = \pm N_c$ carry unit baryon numbers and spin $N_c/2$, corresponding to one-flavor baryons. For an observer on the domain wall, the nature of particle statistics is intrinsically colorful.

\subsection{Baryons as Vortices for $N_f>1$}

Next, we extend the preceding discussion from the single-flavor case to the multi-flavor scenario. In this context, based on the level-rank duality presented in Equation~(\ref{dualityNf}), the flavor symmetry group $U(N_f)$ is promoted to a gauge symmetry. Consequently, the gauge field mediates both quark number and isospin charge. Assuming the minimal coupling remains valid, we can construct a Lagrangian analogous to $\mathcal{L}_A$, expressed as

\begin{equation}
	\mathcal{L}_C[\bm{\phi},\mathbb{A}]= |\partial_\mu\bm{\phi}-\mathrm{i}\mathbb{A}_\mu\bm{\phi}|^2-V_C(\bm{\phi}^\dagger\bm{\phi})+\frac{N_c}{4\pi}\epsilon^{\mu\nu\rho}\operatorname{Tr}\left( \mathbb{A}_\mu\partial_\nu \mathbb{A}_\rho-i\frac{2}{3} \mathbb{A}_\mu \mathbb{A}_\nu \mathbb{A}_\rho\right),
	\label{LC}
\end{equation}
where $\bm{\phi}=(\phi^1,\phi^2,\dots,\phi^{N_f})^T$ denotes the $N_f$ quark fields, and the gauge fields $\mathbb{A}_\mu$ belong to the algebra $\mathfrak{u}(N_f)$.

To simplify, we focus on the case $N_f=2$, where $\mathbb{A}_\mu$ is represented in the adjoint representation as $\mathbb{A}_\mu= \mathbb{A}_\mu^a t^a,\,a=0,1,2,3$, and $t^a=(1/2,\bm{\sigma}/2)$, with $\bm{\sigma}$ being the Pauli matrices. The potential, respecting the $U(2)$ symmetry, is written as
\begin{equation}
	V_C(\bm{\phi}^\dagger\bm{\phi}) = N_c\sum_{I=1}^{\infty}c_{I} \left(\frac{\bm{\phi}^{\dagger}\bm{\phi}}{N_c}-v^2\right)^{I}.
\end{equation}
We hypothesize the existence of vortex solutions in this framework. Consistent with the above analyses, evaluating the topological charge and the statistical spin of the vortices demands scrutiny of their asymptotic behavior. The potential determines the vacuum configuration for $\bm{\phi}$. For simplicity, we choose the vacuum state as $\bm{\phi}_0=\sqrt{N_c}(v_1 e^{in_1 \theta},v_2 e^{in_2 \theta+i\varphi}),\, n_1,n_2\in \mathbb{Z},\, v_1^2+v_2^2=v^2,\, v_1,v_2>0$, where $\varphi$ is an undetermined phase angle. Assuming the vortex is centered at the origin, the asymptotic behavior becomes $\bm{\phi}(r\rightarrow \infty, \theta)=\bm{\phi}_0$. To ensure finite vortex energy, the following condition should hold
\begin{equation}
	\int \mathrm{d}\bm{r}^2|\partial_i\bm{\phi}-i\mathbb{A}_i\bm{\phi}|^2 < \infty. 
\end{equation}

Expanding terms up to $O(\frac{1}{r})$, constraints on the gauge field at large distances can be derived as
\begin{equation}
	\mathbb{A}_i^0(r\rightarrow \infty, \theta)=(n_1+n_2)\frac{\mathbf{e}_\theta}{r}, \quad \mathbb{A}_i^3(r\rightarrow \infty, \theta)=(n_1-n_2)\frac{\mathbf{e}_\theta}{r}.
\end{equation}
For the components $\mathbb{A}_\mu^1$ and $\mathbb{A}_\mu^2$, the analysis differs. Since the vacuum state $\bm{\phi}_0$ can impart mass to three of the four gauge fields, the massless gauge field must vanish in vortex configurations. By substituting $\bm{\phi}_0$ into $\mathcal{L}_C$, the interaction terms involving $\mathbb{A}_\mu^1$ and $\mathbb{A}_\mu^2$ become
\begin{equation}
	\begin{aligned}
		\mathcal{L}_{\mathbb{A}^2} &= \frac{1}{4}(v_1^2+v_2^2)\left[ (\mathbb{A}_\mu^0)^2+(\mathbb{A}_\mu^1)^2+(\mathbb{A}_\mu^2)^2+(\mathbb{A}_\mu^3)^2\right] +\frac{1}{2}(v_1^2-v_2^2)\mathbb{A}_\mu^0 \mathbb{A}_\mu^3 \\ 
		&\quad +v_1 v_2\mathbb{A}_\mu^0\left\{\cos[(n_1-n_2)\theta-\varphi] \mathbb{A}_\mu^1-\sin[(n_1-n_2)\theta-\varphi] \mathbb{A}_\mu^2 \right\}.
	\end{aligned}
\end{equation}
The dependence on the polar angle $\theta$ appears, complicating the definition of mass eigenstates globally and disrupting the vortex’s central symmetry. A straightforward resolution is to set $\mathbb{A}_\mu^1$ and $\mathbb{A}_\mu^2$ to zero everywhere. Under this simplification, the remaining components of $\mathbb{A}_\mu$ are diagonalized
\begin{equation}
	\mathbb{A}_0(r\rightarrow \infty, \theta)=\begin{pmatrix}
		O(\frac{1}{r}) & 0 \\ 
		0 & O(\frac{1}{r})
	\end{pmatrix},\quad
	\mathbb{A}_i(r\rightarrow \infty, \theta)=\begin{pmatrix}
		n_1\frac{\mathbf{e}_\theta}{r}+O(\frac{1}{r}) & 0 \\ 
		0 & n_2\frac{\mathbf{e}_\theta}{r}+O(\frac{1}{r})
	\end{pmatrix},
\end{equation}
where the residual $O(\frac{1}{r})$ term decays exponentially, reflecting the massiveness of $\mathbb{A}_\mu^0$ and $\mathbb{A}_\mu^3$.

With non-zero values of $\mathbb{A}_\mu^0$ and $\mathbb{A}_\mu^3$, the Lagrangian density takes the form
\begin{equation}
	\begin{aligned}
		\mathcal{L}_C[\bm{\phi},\mathbb{A}] &= \left|\partial_\mu\phi_1-i\frac{1}{2}(\mathbb{A}_\mu^0+\mathbb{A}_\mu^3)\phi_1\right|^2+\left|\partial_\mu\phi_2-i\frac{1}{2}(\mathbb{A}_\mu^0-\mathbb{A}_\mu^3)\phi_2\right|^2 -V_C(\bm{\phi}^\dagger\bm{\phi}) \\ 
		&\quad +\frac{N_c}{4\pi}\epsilon^{\mu\nu\rho}\operatorname{Tr}\left( \mathbb{A}_\mu\partial_\nu \mathbb{A}_\rho-i\frac{2}{3} \mathbb{A}_\mu \mathbb{A}_\nu \mathbb{A}_\rho\right),
	\end{aligned}
\end{equation}
where, in this specific two-scalar-field case (since we consider $N_f=2$ here), $\mathbb{A}_\mu^+=(\mathbb{A}_\mu^0+\mathbb{A}_\mu^3)/2$ interacts with $\phi_1$ to propagate a unit quark number, while $\mathbb{A}_\mu^-=(\mathbb{A}_\mu^0-\mathbb{A}_\mu^3)/2$ interacts with $\phi_2$ and similarly propagates a unit quark number. This configuration reduces the complexity of calculations, effectively behaving as two independent copies of the single-flavor case. The gauge field can thus be expressed as
\begin{equation}
	\mathbb{A}_\mu=\begin{pmatrix}
		\mathbb{A}^+_\mu & 0 \\ 
		0 & \mathbb{A}^-_\mu
	\end{pmatrix},\quad
	\mathbb{A}_\mu^{+,-}(r\rightarrow \infty, \theta)=\left(O\!\left(\frac{1}{r}\right),n_{1,2}\frac{\mathbf{e}_\theta}{r}+O\!\left(\frac{1}{r}\right)\right).
\end{equation}

The non-Abelian Chern--Simons term generates a current given by
\begin{equation}
	J^{\mu,a} = \frac{N_c}{4\pi}\epsilon^{\mu\nu\rho}\partial_\nu \mathbb{A}_\rho^a = \frac{N_c}{8\pi}\epsilon^{\mu\nu\rho} \mathbb{F}_{\nu\rho}^a, \quad a = 0, 1, 2, 3,
\end{equation}
where $\mathbb{F}_{\mu\nu}$ represents the field strength tensor of $\mathbb{A}_\mu$. Due to the central symmetry of the vortex solutions, which ensures $[\mathbb{A}_\mu, \mathbb{A}_\nu] = 0$, the currents for $\mathbb{A}_\mu^\pm$ simplify to
\begin{equation}
	J^{\mu,\pm} = J^{\mu,0} \pm J^{\mu,3} = \frac{N_c}{2\pi}\epsilon^{\mu\nu\rho}\partial_\nu \mathbb{A}_\rho^\pm.
\end{equation}
This allows us to compute the fluxes associated with the vortex
\begin{equation}
	\Phi^+ = \int \epsilon^{0\nu\rho}\partial_\nu \mathbb{A}_\rho^+ \, dx\,dy = 2\pi n_1,\quad \Phi^- = \int \epsilon^{0\nu\rho}\partial_\nu \mathbb{A}_\rho^- \, dx\,dy = 2\pi n_2,
\end{equation}
and the corresponding topological charges
\begin{equation}
	Q^+ = \int J^{0,+} \, dx\,dy = n_1 N_c, \quad Q^- = \int J^{0,-} \, dx\,dy = n_2 N_c.
\end{equation}

When two vortices are far apart and weakly interacting, their individual fluxes and charges remain independent. Consequently, the total spin, arising from the combined contributions of fluxes and charges, is additive
\begin{equation}
	S = \frac{\Phi^+ Q^+}{4\pi} + \frac{\Phi^- Q^-}{4\pi} = (n_1^2 + n_2^2) \frac{N_c}{2}, \quad n_1, n_2 \in \mathbb{Z}.
\end{equation}
The baryon number is defined analogously to the single-flavor case as
\begin{equation}
	B = \frac{Q^+ + Q^-}{N_c} = n_1 + n_2.
\end{equation}
Clearly, the baryon number corresponds to the sum of the winding numbers of the individual flavors, with each flavor contributing independently. For instance, setting $n_1=1$ and $n_2=0$ replicates the one-flavor scenario, though $\phi_2 \neq 0$ still contributes to the vortex configuration. Conversely, with $n_1=n_2=1$, the vortex exhibits a baryon number of two and a spin of $N_c$, characteristic of a two-baryon system. Alternatively, choosing $n_1=1$ and $n_2=-1$ yields a baryon-antibaryon configuration, possessing zero baryon number but a spin of $N_c$. Thus, vortices involving multiple flavors inevitably exhibit high spins, with each baryon or antibaryon contributing a minimum spin of $N_c/2$. 

It is worth noting that vortex solutions with multiple flavors inherently carry high spins, aligning with suggestions that high-spin baryons are better described using Hall droplets or vortex configurations. In contrast, the Skyrme model~\cite{Komargodski:2018odf} remains more suitable for describing low-spin baryons. Additionally, $\mathcal{L}_C$ captures only a specific type of vortex configuration. Other potential vortex structures, which could display distinct properties, warrant further investigation in future studies.

In the two-flavor situation discussed above, we considered a vortex configuration with $\mathbb{A}_\mu^0 \neq 0$, $\mathbb{A}_\mu^3 \neq 0$, and $\mathbb{A}_\mu^1 = \mathbb{A}_\mu^2 = 0$, which involves the gauge isospin symmetry. However, if we impose that $\mathbb{A}_\mu^0$ is the only non-zero gauge field component, the Lagrangian $\mathcal{L}_C$ simplifies to
\begin{equation}
	\mathcal{L}_C[\bm{\phi},A] = |\partial_\mu \bm{\phi} - \mathrm{i} C_\mu \bm{\phi}|^2 - V_C(\bm{\phi}^\dagger \bm{\phi}) + \frac{N_c}{2\pi} \epsilon^{\mu\nu\rho} C_\mu \partial_\nu C_\rho,
	\label{LC1}
\end{equation}
where we rescaled $C_\mu = \frac{1}{2}\mathbb{A}_\mu^0$ to normalize the coupling between $C_\mu$ and $\bm{\phi}$ to unity. The Higgs-type potential imposes a non-zero vacuum expectation value, $\bm{\phi}^\dagger \bm{\phi} = N_c v^2 > 0$, leading to a theory that corresponds to a $CP^1 \cong SU(2) / U(1) \times U(1)$ model with an additional Chern--Simons term. 

The equation of motion for $C^\mu$ is given by
\begin{equation}
	\mathrm{i} (\bm{\phi}^\dagger \partial^\mu \bm{\phi} - \bm{\phi} \partial^\mu \bm{\phi}^\dagger) + 2 C^\mu \bm{\phi}^\dagger \bm{\phi} + \frac{N_c}{\pi} \epsilon^{\mu\nu\rho} \partial_\nu C_\rho = 0.
\end{equation}
Neglecting the dynamical effects of the Chern--Simons term, we can approximate
\begin{equation}
	C^\mu \simeq -\frac{\mathrm{i}}{2 \bm{\phi}^\dagger \bm{\phi}} (\bm{\phi}^\dagger \partial^\mu \bm{\phi} - \bm{\phi} \partial^\mu \bm{\phi}^\dagger).
\end{equation}

Regardless of the specific value of $v$, we can rescale $\bm{\phi} \rightarrow v \bm{\phi}$ to normalize $\bm{\phi}^\dagger \bm{\phi} = N_c$. This results in
\begin{equation}
	C^\mu = -\frac{\mathrm{i}}{2} \left(\bm{\phi}_c^\dagger \partial^\mu \bm{\phi}_c - \bm{\phi}_c^\dagger \partial^\mu \bm{\phi}_c \right), \quad \text{with} \quad \bm{\phi}_c = \frac{\bm{\phi}}{\sqrt{N_c}}.
\end{equation}

Introducing the parameter $m^i = \bm{\phi}^\dagger \sigma^i \bm{\phi}$ with $i = 1, 2, 3$ and $\bm{m}^2 = 1$, and considering a stable finite-density system with quark chemical potential $C^0 = 2\mu_q$ (the factor of 2 arises from two quark flavors), the Chern--Simons term becomes
\begin{equation}
	\frac{N_c}{2\pi} \epsilon^{\mu\nu\rho} C_\mu \partial_\nu C_\rho \simeq -N_c \mu_q \frac{\mathrm{i}}{2\pi} \epsilon^{ij} \partial_i \bm{\phi}^\dagger \partial_j \bm{\phi} = N_c \mu_q \frac{1}{8\pi} \epsilon^{ij} \bm{m} \cdot (\partial_i \bm{m} \times \partial_j \bm{m}),
\end{equation}
showing explicitly that the Chern--Simons term on the $\eta'$ domain wall reduces to the Wess--Zumino--Witten (WZW) term. At low energy, the vector field can be integrated out, leaving an effective $CP^1$ model with a WZW term. For general flavors $N_f \geq 2$, the effective theory generalizes to a $CP^{N_f-1}$ model. The $CP^{N_f-1}$ model inherently possesses a hidden global $U(1)$ symmetry, which allows for the introduction of an $A^\mu$ field by gauging this symmetry.

A noteworthy phenomenon is the emergence of a domain wall Skyrmion phase under rotation~\cite{Eto:2023tuu,Eto:2023rzd}, where domain walls consist of $\eta^\prime$ and $\pi^0$ meson fields. In the two-flavor case, the conventional $(3+1)$-dimensional Skyrmions flow into $(2+1)$-dimensional domain walls, transforming into baby skyrmions~\cite{Kudryavtsev:1996er,Battye:2013tka,Winyard:2015dba,Leask:2024ith}. The induced effective theory on the domain wall is a $CP^1$ model with a WZW term. Thus, our vortex picture for baryons partially aligns with the domain wall Skyrmion description. Additionally, under a magnetic field, domain-wall skyrmion chains can emerge~\cite{Bigazzi:2022luo,Eto:2023tuu,Eto:2023rzd,Huang:2017pqe,Huang:2019rkz,Nishimura:2020odq,Eto:2021gyy,Nitta:2012wi,Eto:2015uqa,Gudnason:2014nba,Nitta:2022ahj,Eto:2023lyo,Eto:2023wul,Qiu:2024zpg,Chen:2021vou,Fukushima:2018ohd}.

\section{Topological Chiral Bag Model} \label{sec5}

Baryons are successfully constructed as vortices on the $(2+1)$ dimensional $\eta'$ domain wall, with the flux of the $U(1)$ Chern--Simons field serving as a measure of the baryon number. A natural question arises: what is the source of the flux in $(3+1)$ dimensions? Such sources should interact with the $\eta'$ field and induce the corresponding baryon number. In Section~\ref{subchiral}, we reviewed the chiral bag model with the $\eta'$ boundary condition, where the color charge flow-out effect was identified and subsequently resolved. Moreover, the $\eta'$ field experiences a sharp discontinuity at the bag surface, effectively rendering the boundary a domain wall. Thus, our investigation of the $\eta'$ domain wall can be applied to enhance the understanding of the chiral bag model.

Since the bag boundary behaves as a domain wall, it is reasonable to hypothesize that the dynamics of quarks confined within the bag induce baryon number on the bag surface. In fact, the quarks confined inside the bag act analogously to monopoles, with their flux inducing baryon numbers on the $\eta'$ boundary~\cite{Komargodski:2018odf,Ma:2019xtx,Lin:2023qya,Karasik:2020zyo,Karasik:2020pwu}. The outflow of color charge is accompanied by that of baryon number, requiring the introduction of a counterterm to preserve one-form gauge invariance~\cite{Dierigl:2014xta}.

\subsection{Confined Monopoles Inside the Chiral Bag}

The core idea of the chiral bag model is to impose a boundary condition that not only confines quarks and gluons within the bag but also connects to the dynamics of the mesons outside. If we only consider the pseudoscalar isosinglet $\eta'$ meson field, the boundary essentially becomes a domain wall. The dynamics on the bag surface should be induced by the physics inside the bag, making the confined quarks essential for a complete description.

To describe a color singlet formed by quarks inside the bag, a mechanism that causes confinement is necessary. It is worth noting that in Equation~\eqref{dQ}, color charge is induced when a non-zero color magnetic field crosses the surface with a constant $\eta'$. One can further conjecture the existence of monopoles inside the bag, which generate significant color magnetic fields on the bag surface. Since gluons are governed by the $SU(N_c)$ Yang--Mills theory, quarks inside the bag can be modeled as monopoles, whose condensation leads to confinement. It is suggested that the classification of monopole charges follows the discrete group $Z_{N_c}$, which is the center of the gauge group $SU(N_c)$. The topological properties of $SU(N_c)$ gauge theory can then be described within the framework of topological field theory~\cite{Dierigl:2014xta,Banks:2010zn,Gukov:2013zka}. Using a magnetic Abelian Higgs model, the discrete gauge group $Z_{N_c}$ can be embedded in a $U(1)$ theory.

To describe the monopoles confined within the bag, we can follow the method proposed in~\cite{Dierigl:2014xta}. Consider a complex scalar field $\Phi$ that carries $N_c$ magnetic charges, with $\tilde{A}$ representing the magnetic dual gauge field of the usual $U(1)$ gauge field. The gauge covariant derivative is expressed as
\begin{equation}
	D\Phi=(\mathrm{d}-iN_c\tilde{A})\Phi.
\end{equation}
The %MDPI: 
complex scalar field $\Phi$ represents the monopole density and is assumed to have a Higgs-type potential, resulting in a non-zero vacuum expectation value $\left\langle\left|\Phi\right|\right\rangle\equiv \upsilon > 0$. In the low-energy region, $\Phi=\upsilon e^{i\varphi}$, and the action is dominated by pure gauge configurations satisfying
\begin{equation}
	D\Phi=i(\mathrm{d}\varphi-N_c\tilde{A})\upsilon \quad\Rightarrow\quad \mathrm{d}\varphi-N_c\tilde{A}=0,
\end{equation}
which precisely describes the discrete $Z_{N_c}$ gauge theory. By dualizing the magnetic field $\tilde{A}$ back to the electric description via the usual gauge field $A$, one obtains the action of a BF theory~\cite{Horowitz:1989ng}
\begin{equation}
	S=\frac{i}{2\pi} \int \mathrm{d}\tilde{A} \wedge (\mathrm{d}A - N_c B).
\end{equation}
This action possesses a 1-form gauge symmetry parametrized by a 1-form $\lambda$, which satisfies the quantization condition over a closed 2-surface, in this case, the bag surface $\Sigma$
\begin{equation}
	\frac{1}{2\pi}\oint \mathrm{d}\lambda\in\mathbb{Z}, \quad A \to A + N_c\lambda, \quad B \to B + \mathrm{d}\lambda.
\end{equation}
This simplies that the magnetic charge $\Delta m$ added into the bag must be a multiple of the condensed monopole charge $N_c$
\begin{equation}
	m=\frac{1}{2\pi}\oint \mathrm{d}A \to m + \frac{1}{2\pi}\oint N_c \mathrm{d}\lambda = m + jN_c, \quad j\in\mathbb{Z}.
	\label{Char}
\end{equation}

Now, let us consider what happens when the $\eta^\prime$ field crosses the bag surface. This situation arises due to the transformation similarity between the $\eta^\prime$ field and $\theta$, leading to vacuum transitions between different branches. As shown in Equation~\eqref{Leta}, the $\eta^\prime$ and $\theta$ fields together exhibit multiple vacua labeled by \( n \), resulting in the presence of domain walls. In the \( SU(N_c) \) Yang--Mills theory with a \( \theta \)-term, 't Hooft anomaly matching dictates that each vacuum label \( n \) is associated with a topological term in the action~\cite{Kitano:2020evx}
\begin{equation}
	S = \frac{i}{2\pi} \int d\tilde{A} \wedge (dA - N_c B) + \frac{N_c \theta}{4\pi} B \wedge B - \frac{N_c n}{2} B \wedge B.
\end{equation}	
To derive the interaction between $\eta^\prime$ and the gauge fields \( A \) and \( B \), we shift \( \theta \to \theta + N_f \eta^\prime \) and then set \( \theta = 0 \). The action becomes
\begin{equation}
	S = \frac{i}{2\pi} \int d\tilde{A} \wedge (dA - N_c B) + \frac{N_c N_f \eta^\prime}{4\pi} B \wedge B - \frac{N_c n}{2} B \wedge B.
\end{equation}
By naively setting \( \eta_{\mathrm{in}}^\prime = 0 \) inside the bag and \( \eta_{\Sigma}^\prime = 2\pi \) outside, the bag boundary effectively behaves as an $\eta^\prime$ domain wall, where \( N_f \eta^\prime \) undergoes a \( 2\pi N_f \) shift. Since a shift of \( \Delta \eta^\prime = 2\pi \) corresponds to a topological shift of \( \Delta n = N_f \), we interpret the bag surface as a domain wall where vacuum branches transition with \( \Delta n = N_f \). Consequently, the bag surface separates two different vacua, and the effective action is not invariant under 1-form gauge transformation
\begin{equation}
	\Delta S=-\frac{i}{2\pi}\int \mathrm{d}\left[n \lambda\wedge \mathrm{d}A+\frac{nN_c}{2}\lambda\wedge \mathrm{d}\lambda\right]+\frac{iN_c}{4\pi}\int \mathrm{d}n\wedge(2\lambda\wedge B+\lambda\wedge \mathrm{d}\lambda).
\end{equation}
The first term is a total derivative and does not contribute here since our bag surface is a closed surface. However, the second term develops a contribution on the bag surface
\begin{equation}
	\Delta S_{\mathrm{surface}}=-\frac{iN_c}{4\pi}\int_{\Sigma}(2\lambda\wedge B+\lambda\wedge \mathrm{d}\lambda).
\end{equation}
To maintain gauge invariance, the dynamics on the bag surface must be non-trivial. Based on arguments from $\mathcal{N} = 1$ supersymmetric Yang--Mills theory~\cite{Gaiotto:2017yup}, it is suggested that on the bag surface, there exists an $SU(N_c)_{N_f}$ Chern--Simons theory. Due to level-rank duality, an $SU(N_c)_{N_f}$ Chern--Simons theory is identified with a $U(N_f)_{-N_c}$ Chern--Simons theory, leading to the effective action proposed in~\cite{Kitano:2020evx}
\begin{equation}
	S_\Sigma= -\frac{i}{4\pi}\left[N_c\operatorname{tr}\left(\mathbb{A}\mathrm{d}\mathbb{A}-i\frac{2}{3}\mathbb{A}^3\right) + 2\operatorname{tr}(\mathbb{A})\mathrm{d}A\right], \label{eff1}
\end{equation}
where $\mathbb{A}$ is a 1-form $U(N_f)$ gauge field that transforms as
\begin{equation}
	\mathbb{A}\rightarrow \mathbb{A} - \lambda \bm{1}_{N_f \times N_f}.
\end{equation}
It is straightforward to verify that, under the 1-form gauge transformation, the contribution from the bag surface $S_\Sigma$ exactly offsets $\Delta S_{\mathrm{surface}}$.

To investigate the dynamics on the bag surface for the topological field, we observe that $A$ couples with $\mathbb{A}$, behaving like a background field for the dynamical bag surface. Explicitly applying Gauss's law on the bag surface yields
\begin{equation}
	\mathrm{d}\mathbb{A}-i\mathbb{A}^2=-\frac{\mathrm{d}A}{N_c}\bm{1}_{N_f \times N_f}.
\end{equation}
Thus, the vector field $\mathbb{A}$ acquires magnetic flux
\begin{equation}
	\frac{1}{2\pi N_f}\int_\Sigma \operatorname{tr}(\mathrm{d}\mathbb{A})=-\frac{1}{2\pi N_c}\int_\Sigma \mathrm{d}A,
	\label{Flux}
\end{equation}
indicating that only the $U(1)$ part of $\mathbb{A}$ acquires flux. Since $\mathbb{A}$ describes the monopole density inside the bag in a topological manner, we can restrict the complex scalar field $\Phi$ on the bag surface as $\phi=\Phi|_\Omega$. The field $\phi$ inherits a Higgs-type potential $V(\phi^*\phi)$ from $\Phi$ and minimally couples with $\mathbb{A}$. For the one-flavor case, we can write down an effective theory at leading order on the bag surface
\begin{equation}
	\int_\Sigma|\mathrm{d}\phi-\mathrm{i}a\phi|^2-\frac{N_c}{4\pi}a\mathrm{d}a-V(\phi^*\phi).
\end{equation}
where we use $a$ to replace $\mathbb{A}$ for the one-flavor case. This is precisely the Chern--Simons--Higgs theory conjectured to exist on the $\eta^\prime$ domain wall, as described in Equation~\eqref{LA}, with vortex solutions proposed to represent baryons or multi-baryon structures in $(2+1)$ dimensions. For the multi-flavor case, the theory generalizes to a non-Abelian one
\begin{equation}
	\int_\Sigma|\mathrm{d}\bm{\phi}-\mathrm{i}\mathbb{A}\bm{\phi}|^2-\frac{N_c}{4\pi}\operatorname{tr}\left(\mathbb{A}\mathrm{d}\mathbb{A}-i\frac{2}{3}\mathbb{A}^3\right)-V(\bm{\phi}^\dagger\bm{\phi}),
\end{equation}
where $\bm{\phi}=(\phi^1,\phi^2,\ldots,\phi^{N_f})^T$ is a complex field with $N_f$ components, the same as Equation~\eqref{LB}.

Thus, we see that there indeed exists a duality for the Chern--Simons theory shown in Equation~\eqref{LC} on the $\eta^\prime$ domain wall~\cite{Gaiotto:2017yup,Gaiotto:2017tne}. Under level-rank duality, the dual $SU(N_c)_{N_f}$ Chern--Simons theory emerges on the bag surface. Notably, the counterterm added to seal off the color leak, shown in Equation~\eqref{CTCS}, is also expressed as an $SU(N_c)_{N_f}$ Chern--Simons theory coupled to $\eta^\prime$. A bold but natural conjecture is that these two Chern--Simons theories are identified, providing deeper insight into the chiral bag model.

\subsection{Block the Outflow of Color Charge }

We have seen that if confinement is assumed to be caused by monopoles, one can place the monopoles inside the bag. The topological properties can be described by a $\mathbb{Z}_{N_c}$ gauge field $A$, whose flux measures the quantity of confined monopoles. On the bag surface, a $U(N_f)_{-N_c}$ Chern--Simons theory is induced concerning the new dynamical field $\mathbb{A}$, which also carries non-zero flux. However, as shown in Equation~\eqref{Flux}, the corresponding flux quantity has an opposite sign, indicating that the bag surface should contain anti-monopoles. This arrangement effectively seals off the color leak for the chiral bag model, as shown in Equation~\eqref{CTCS}.

Now, let us focus on the quantities of monopoles inside the bag and on the bag surface. In the previous section, we chose $\eta_{\mathrm{in}}' = 0$ inside the bag and $\eta_{\Sigma}' = 2\pi$ outside. For a more general bag model boundary condition, we consider both $\eta_{\Sigma}'$ and $\eta_{\mathrm{in}}'$ to be arbitrary. The effective theory on the bag surface, as shown in Equation~\eqref{eff1}, should be modified as follows:
\begin{equation}
	S_\Sigma= -\left(\frac{\eta_{\Sigma}'-\eta_{\mathrm{in}}'}{2\pi}\right) \times \frac{i}{4\pi}\left[N_c\operatorname{tr}\left(\mathbb{A}\mathrm{d}\mathbb{A}-i\frac{2}{3}\mathbb{A}^3\right) + 2\operatorname{tr}(\mathbb{A})\mathrm{d}A\right],
	\label{CS}
\end{equation}
where the flux relation from Gauss's law, Equation~\eqref{Flux}, remains unchanged, but the shift in $\eta'$ introduces the Witten effect~\cite{Witten:1979ey}. The monopoles inside the bag carry the magnetic field $B_A = \mathrm{d}A/N_c$, and when this passes through the bag surface, $B_A$ interacts with $\eta'$, generating the electric charge $Q_A$
\begin{equation}
	\frac{\mathrm{d}Q_A}{\mathrm{d}\eta^\prime_\Sigma}=\frac{\mathrm{d}}{\mathrm{d}\eta^\prime_\Sigma}\oint_{\Sigma}\mathrm{d}S\boldsymbol{E}_A\cdot\bm{n}=\frac{1}{8\pi^2}\oint_\Sigma\mathrm{d}S\boldsymbol{B}_A\cdot\bm{n}, \label{leak1}
\end{equation}
where we have implicitly fixed $\eta_{\mathrm{in}}'$.

Meanwhile, the dynamical field $\mathbb{A}$ on the bag surface also possesses non-zero flux due to Gauss's law, leading to the corresponding electric charge $Q_\mathbb{A}$
\begin{equation}
	\frac{\mathrm{d}Q_\mathbb{A}}{\mathrm{d}\eta^\prime_\Sigma}=\frac{\mathrm{d}}{\mathrm{d}\eta^\prime_\Sigma}\oint_{\Sigma}\mathrm{d}S\boldsymbol{E}_\mathbb{A}\cdot\bm{n}=\frac{1}{8\pi^2}\oint_\Sigma\mathrm{d}S\boldsymbol{B}_\mathbb{A}\cdot\bm{n},
\end{equation}
where $B_\mathbb{A}=\operatorname{tr}(\mathrm{d}\mathbb{A})/N_f$. According to the flux relation in Equation~\eqref{Flux}, we find
\begin{equation}
	\frac{\mathrm{d}Q_A}{\mathrm{d}\eta^\prime_\Sigma}+\frac{\mathrm{d}Q_\mathbb{A}}{\mathrm{d}\eta^\prime_\Sigma}=0.
\end{equation}

We observe that $B_\mathbb{A}$ corresponds to the $U(1)$ part in flavor space associated with baryon number conservation symmetry, so the electric charges $Q_A$ and $Q_\mathbb{A}$ both correspond to quark number or baryon number. Therefore, we have a non-zero $\mathrm{d}Q_A$ baryon number flowing out from the bag once $\eta'_\Sigma$ is time-dependent, similar to the color charge leaking reviewed in Section~\ref{subchiral}. Fortunately, the topological field theory on the bag surface contributes the same quantity but with opposite sign for the baryon charge, $\mathrm{d}Q_{\mathbb{A}}$, effectively blocking the leak!

In fact, the dynamical field $\mathbb{A}$ on the bag surface was originally proposed to preserve gauge invariance, which necessarily implies charge conservation. On the other hand, multiples of $N_c$ monopoles condense to cause confinement and must carry color charge. Note that, here monopoles behave similarly to quarks, but we do not identify them explicitly. The nature of the baryon number carriers remains unclear; both quarks and gluons are possible, and all baryon numbers we consider are induced. The leakage of baryon number is accompanied by color charge leakage, and we can replace the baryon number charge with the color charge in Equation~\eqref{leak1}
\begin{equation}
	\frac{\mathrm{d}Q_G^a}{\mathrm{d}\eta^\prime_\Sigma} = \frac{\mathrm{d}}{\mathrm{d}\eta^\prime_\Sigma}\oint_{\Sigma}\mathrm{d}S\boldsymbol{E}_G^a\cdot\bm{n} = \frac{1}{8\pi^2}\oint_\Sigma\mathrm{d}S\boldsymbol{B}_G^a\cdot\bm{n}, \label{leak1}
\end{equation}
which matches the color charge leakage found in Equation~\eqref{dQ} after absorbing the coupling constant $g$. Similarly, the Chern--Simons field theory $U(N_f)_{-N_c}$ on the bag surface, dual to $SU(N_c)_{N_f}$, contributes the same quantity but with opposite sign for the color charge, $-\mathrm{d}Q_G^a$, which precisely cancels the color charge leakage. Thus, the Chern--Simons theory in Equation~\eqref{CS}, initially introduced to maintain gauge invariance, also acts as a counterterm, as seen in Equation~\eqref{CTCS}. Under level-rank duality, these two forms of topological theory describe the same physical phenomenon from different perspectives.

We have seen that, on the one hand, the $SU(N_c)_{N_f}$ Chern--Simons theory on the bag surface blocks the color charge leakage, restoring gauge invariance and ensuring confinement. On the other hand, the dual $U(N_f)_{-N_c}$ Chern--Simons theory introduces quark number, with the same magnitude but opposite sign relative to the quarks inside the bag. Combining these contributions yields a chiral bag model for baryons with zero net baryon number. Thus, an additional mechanism is required to generate baryon number in the chiral bag model.

If we set $\eta_{\mathrm{in}}' = 0$ inside the bag and $\eta_{\Sigma}' = 2\pi$ outside, the theory on the bag surface becomes independent of the choice of the interior configuration, as shown in~\cite{Kitano:2020evx}. Additional fields can also exist on the bag surface. While we have focused on the chiral bag's interior, where quarks are represented by monopoles, the exterior of the bag is equally interesting as it hosts meson fields acting as background fields.

On the bag surface, a dynamical $U(N_f)$ vector field exists, suggesting that the vector meson fields in flavor space behave similarly. The vector meson field $V$ interacts with the dynamical field $\mathbb{A}$ on the bag surface~\cite{Kitano:2020evx}
%\begin{adjustwidth}{-\extralength}{0cm}
	\be
%		\begin{array}{lll}
			S_\Sigma & = &{} -\left(\frac{\eta_{\Sigma}' - \eta_{\mathrm{in}}'}{2\pi}\right) \\ 
			& &{} \times \frac{i}{4\pi}\int_{\Sigma}\left[N_c\operatorname{tr}\left(\mathbb{A}\mathrm{d}\mathbb{A}-i\frac{2}{3}\mathbb{A}^3\right) - N_c\operatorname{tr}\left(V\mathrm{d}V - i\frac{2}{3}V^3\right) + 2(\operatorname{tr}(V) + \operatorname{tr}(\mathbb{A}))\mathrm{d}A\right]. \nonumber
%		\end{array}
	\ee
%\end{adjustwidth}
The vector meson field transforms as
\begin{equation}
	V \rightarrow V - \lambda \bm{1}_{N_f \times N_f}.
\end{equation}
Thus, the action is invariant under both 1-form and 0-form gauge transformations.

Applying Gauss's law to $V$, we derive a flux relation similar to Equation~\eqref{Flux}
\begin{equation}
	\frac{1}{2\pi}\int_\Sigma \operatorname{tr}(\mathrm{d}V) = -\frac{N_f}{2\pi N_c}\int_\Sigma \mathrm{d}A.
\end{equation}
Focusing on the $U(1)_B$ part of the vector meson field, $V_B = \operatorname{tr}(V)/N_f$, we find that the Witten effect on $V$ gives the baryon number $Q_\mathrm{in}$
\begin{equation}
	Q_\mathrm{in} = -\left(\frac{\eta_{\Sigma}' - \eta_{\mathrm{in}}'}{2\pi}\right) \times \frac{1}{2\pi}\int_\Sigma \operatorname{tr}(\mathrm{d}V) = N_f\left(\frac{\eta_{\Sigma}' - \eta_{\mathrm{in}}'}{2\pi}\right) \times \frac{1}{2\pi N_c}\int_\Sigma \mathrm{d}A.
\end{equation}
For simplicity, consider $N_f = 1$, $\eta_{\mathrm{in}}' = 0$, $\eta_{\Sigma}' = 2\pi$, and $\frac{1}{2\pi}\int_\Sigma \mathrm{d}A = N_c$. This setup represents a chiral bag model for a one-flavor baryon. Outside the bag, the flux of the vector meson field $V$, induced on the surface, extends to infinity or terminates on another chiral bag as an anti-baryon.

To link the bag surface theory with the exterior, $\eta'$ must couple to the vector meson. The effective interaction term for the meson field is
\begin{equation}
	i\frac{N_c}{8\pi^2}\int \eta' \operatorname{tr}(\mathrm{d}V \mathrm{d}V).
\end{equation}
This term appears in the hidden Wess--Zumino term~\cite{Harada:1992np,Harada:2003jx}, with related physics discussed in~\cite{Karasik:2022tmd,Karasik:2020zyo,Karasik:2020pwu}. The baryon number outside the bag, $Q_\mathrm{out}$, is proposed as
\begin{equation}
	Q_\mathrm{out} = -\int_\Xi \left(\frac{\mathrm{d}\eta'}{2\pi}\right) \times \frac{1}{2\pi} \mathrm{d}\operatorname{tr}(V),
\end{equation}
where the integral is taken over the space outside the bag, labeled $\Xi$. The total baryon number of the bag is then $Q = Q_\mathrm{in} + Q_\mathrm{out}$.

It is worth noting that in \cite{Karasik:2022tmd,Karasik:2020zyo,Karasik:2020pwu}, the dynamical $U(N_f)_{-N_c}$ vector field $\mathbb{A}$ is directly identified with the vector meson fields. However, this identification is premature. Here, we observe that both $\mathbb{A}$ and $V$ exist on the bag surface and cannot be distinguished based solely on their positions. Nevertheless, only $V$ extends to the bag's exterior and should be identified with the vector meson fields, in terms of the hidden local flavor symmetry~\cite{Bando:1984ej,Bando:1987br,Harada:2003jx}.

On the bag surface, there exist two vector fields, $\mathbb{A}$ and $V$, both carrying flux but with opposite signs. This configuration results in baryon numbers of equal magnitude but opposite sign, effectively canceling each other. This suggests the possible existence of a meson cloud assembled on the bag surface, comprising a quark and an anti-quark, which resembles a dipole structure excited by the monopoles inside the bag. Thus, a baryon can be understood as a monopole surrounded by a meson cloud. If the singularity of the monopoles can be eliminated by certain methods, the meson cloud could be identified as a baryon. Indeed, the soliton structure formed from the pion field, famously known as a skyrmion, can be interpreted as a baryon. Our description of the chiral bag model provides an explanation for why skyrmions are understood as baryons. However, incorporating the pion field requires modifying the chiral bag boundary condition~\cite{Chodos:1974je,Rho:1983bh,Goldstone:1983tu,Dreiner:1988jg}, which introduces additional complexities and warrants further investigation.

\section{Conclusions and Discussion}\label{sec6}

As extensively discussed above, the concept of baryons as topological solitons is widely accepted. For multiple flavors, baryons are described as skyrmions in $(3+1)$ dimensional spacetime, while for a single flavor, baryons manifest as quantum Hall droplets constructed from an extended, meta-stable configuration of the $\eta^\prime$ field in $(2+1)$ dimensions~\cite{Komargodski:2018odf} or as vortices on $(2+1)$ dimensional $\eta^\prime$ domain walls.

The relation between skyrmions and one-flavor baryon construction has been investigated~\cite{Karasik:2020zyo,Karasik:2020pwu}. However, constructing a one-flavor baryon from the soliton perspective in $(3+1)$-dimensional spacetime remains an open problem, requiring further attention. Moreover, while ample evidence suggests that the topological soliton structure of the meson field gives rise to baryons, the precise mechanism by which high-energy QCD flows to the low-energy effective theory remains unclear and warrants further study~\cite{Witten:1983tw,Karasik:2022tmd,Niemi:1984vz,Zahed:1986qz}. This phenomenon appears to be closely tied to chiral symmetry breaking and the generation of Nambu--Goldstone bosons.

The role of vector mesons is particularly significant; under level-rank duality, vector mesons can be interpreted as dual gluons. However, identifying the physical content of the scalar field in strong dynamics remains challenging, as it is in the fundamental representation of $SU(N_f)$. Since the level-rank duality with scalars is associated with an assumed non-trivial infrared fixed point~\cite{Hsin:2016blu}, the dual theory with scalars might have close connections to the recently developed chiral-scale effective theory of strong interactions, which is based on the assumed nonperturbative infrared fixed point of QCD~\cite{Crewther:2013vea,Li:2016uzn}. 

An environment that realizes this relationship/duality is dense nuclear matter, where the vector manifestation point~\cite{Harada:2000kb,Harada:2001rf} may exist, or where chiral symmetry may be realized in a vector mode~\cite{Georgi:1989xy}, such that the vector meson fields become massless. In this case, the scale symmetry locked to the infrared behavior of strong interactions may emerge~\cite{Ma:2023ugl}. These unresolved questions merit further investigation.

\vspace{6pt}
\authorcontributions{Both authors contributed equally to this review.  %MDPI: For research articles with several authors, a short paragraph specifying their individual contributions must be provided. The following statements should be used ``Conceptualization, X.X. and Y.Y.; methodology, X.X.; software, X.X.; validation, X.X., Y.Y. and Z.Z.; formal analysis, X.X.; investigation, X.X.; resources, X.X.; data curation, X.X.; writing---original draft preparation, X.X.; writing---review and editing, X.X.; visualization, X.X.; supervision, X.X.; project administration, X.X.; funding acquisition, Y.Y. All authors have read and agreed to the published version of the manuscript.'', please turn to the  \href{http://img.mdpi.org/data/contributor-role-instruction.pdf}{CRediT taxonomy} for the term explanation. Authorship must be limited to those who have contributed substantially to the work~reported.
}

\funding{The work of Y.~L. M. is supported in part by the National Science Foundation of China (NSFC) under Grant No. 12347103, the National Key R\&D Program of China under Grant No. 2021YFC2202900, and the Gusu Talent Innovation Program under Grant No. ZXL2024363. }
%MDPI:  Please add: This research received no external funding or This research was funded by [name of funder] grant number [xxx] And The APC was funded by [XXX]. Information regarding the funder and the funding number should be provided. Please check the accuracy of funding data and any other information carefully

%\dataavailability{\hl{ } %MDPI: We encourage all authors of articles published in MDPI journals to share their research data. In this section, please provide details regarding where data supporting reported results can be found, including links to publicly archived datasets analyzed or generated during the study. Where no new data were created, or where data is unavailable due to privacy or ethical restrictions, a statement is still required. Suggested Data Availability Statements are available in section ``MDPI Research Data Policies'' at \url{https://www.mdpi.com/ethics}.} 

%\acknowledgments{\hl{The} %MDPI: 
% work of Y.~L. M. is supported in part by the National Science Foundation of China (NSFC) under Grant No. 12347103, the National Key R\&D Program of China under Grant No. 2021YFC2202900 and Gusu Talent Innovation Program under Grant No. ZXL2024363.}
% %MDPI: To AE: please confirm if the funding information in the Acknowledgments Section should be moved to the Funding Section.

\conflictsofinterest{ The authors declare no conflicts of interest.  %MDPI: Declare conflicts of interest or state ``The authors declare no conflicts of interest.'' Authors must identify and declare any personal circumstances or interest that may be perceived as inappropriately influencing the representation or interpretation of reported research results. Any role of the funders in the design of the study; in the collection, analyses or interpretation of data; in the writing of the manuscript; or in the decision to publish the results must be declared in this section. If there is no role, please state ``The funders had no role in the design of the study; in the collection, analyses, or interpretation of data; in the writing of the manuscript; or in the decision to publish the results''.
} 

%%%%%%%%%%%%%%%%%%%%%%%%%%%%%%%%%%%%%%%%%%
\begin{adjustwidth}{-\extralength}{0cm}

\reftitle{References}

\PublishersNote{}
\end{adjustwidth}

\begin{thebibliography}{999}


%\cite{Gross:1973id}
\bibitem{Gross:1973id}
Gross, D.J.; Wilczek, F.
Ultraviolet Behavior of Nonabelian Gauge Theories. %MDPI: we displayed all the title of references, please confirm.
\emph{Phys. Rev. Lett.} \textbf{1973}, \emph{30}, 1343--1346.
\url{https://doi.org/10.1103/PhysRevLett.30.1343}.
%6561 citations counted in INSPIRE as of 21 Jan 2025


%\cite{Politzer:1973fx}
\bibitem{Politzer:1973fx}
Politzer, H.D.
Reliable Perturbative Results for Strong Interactions?
\emph{Phys. Rev. Lett.} \textbf{1973}, \emph{30}, 1346--1349.
\url{https://doi.org/10.1103/PhysRevLett.30.1346}.
%6267 citations counted in INSPIRE as of 21 Jan 2025


%\cite{Marciano:1977su}
\bibitem{Marciano:1977su}
Marciano, W.J.; Pagels, H..
Quantum Chromodynamics: A Review.
\emph{Phys. Rep.} \textbf{1978}, \emph{36}, 137.
\url{https://doi.org/10.1016/0370-1573(78)90208-9}.
%998 citations counted in INSPIRE as of 21 Jan 2025


%\cite{Callan:1969sn}
\bibitem{Callan:1969sn}
Callan, C.G., Jr.; Coleman, S.R.; Wess, J.; Zumino, B.
Structure of phenomenological Lagrangians. 2.
\emph{Phys. Rev.} \textbf{1969}, \emph{177}, 2247--2250.
\url{https://doi.org/10.1103/PhysRev.177.2247}.
%2156 citations counted in INSPIRE as of 31 Dec 2024


%\cite{Coleman:1969sm}
\bibitem{Coleman:1969sm}
Coleman, S.R.; Wess, J.; Zumino, B.
Structure of phenomenological Lagrangians. 1.
\emph{Phys. Rev.} \textbf{1969}, \emph{177}, 2239--2247.
\url{https://doi.org/10.1103/PhysRev.177.2239}.
%2409 citations counted in INSPIRE as of 31 Dec 2024


%\cite{Weinberg:1978kz}
\bibitem{Weinberg:1978kz}
Weinberg, S.
Phenomenological Lagrangians.
\emph{Physica A} \textbf{1979}, \emph{96},  327--340.
\url{https://doi.org/10.1016/0378-4371(79)90223-1}.
%4048 citations counted in INSPIRE as of 31 Dec 2024


%\cite{Gasser:1983yg}
\bibitem{Gasser:1983yg}
Gasser, J.; Leutwyler, H.
Chiral Perturbation Theory to One Loop.
\emph{Ann. Phys.} \textbf{1984}, \emph{158}, 142.
\url{https://doi.org/10.1016/0003-4916(84)90242-2}.
%4817 citations counted in INSPIRE as of 21 Jan 2025


%\cite{Gasser:1984gg}
\bibitem{Gasser:1984gg}
Gasser, J.; Leutwyler, H.
Chiral Perturbation Theory: Expansions in the Mass of the Strange Quark.
\emph{Nucl. Phys. B} \textbf{1985}, \emph{250}, 465--516.
\url{https://doi.org/10.1016/0550-3213(85)90492-4}.
%4408 citations counted in INSPIRE as of 21 Jan 2025



%\cite{Weinberg:1990rz}
\bibitem{Weinberg:1990rz}
Weinberg, S.
Nuclear forces from chiral Lagrangians.
\emph{Phys. Lett. B} \textbf{1990}, \emph{251}, 288--292.
\url{https://doi.org/10.1016/0370-2693(90)90938-3}.
%1653 citations counted in INSPIRE as of 31 Dec 2024


%\cite{Gasser:1987rb}
\bibitem{Gasser:1987rb}
Gasser, J.; Sainio, M.E.; Svarc, A.
Nucleons with chiral loops.
\emph{Nucl. Phys. B} \textbf{1988}, \emph{307}, 779--853.
\url{https://doi.org/10.1016/0550-3213(88)90108-3}.
%1012 citations counted in INSPIRE as of 21 Jan 2025


%\cite{Hatsuda:1994pi}
\bibitem{Hatsuda:1994pi}
Hatsuda, T.; Kunihiro, T.
QCD phenomenology based on a chiral effective Lagrangian.
\emph{Phys. Rep.} \textbf{1994}, \emph{247}, 221--367.
\url{https://doi.org/10.1016/0370-1573(94)90022-1}.
%[arXiv:hep-ph/9401310 [hep-ph]].
%1866 citations counted in INSPIRE as of 21 Jan 2025


%\cite{Bernard:2007zu}
\bibitem{Bernard:2007zu}
Bernard, V.
Chiral Perturbation Theory and Baryon Properties.
\emph{Prog. Part. Nucl. Phys.} \textbf{2008}, \emph{60}, 82--160.
\url{https://doi.org/10.1016/j.ppnp.2007.07.001}.
%[arXiv:0706.0312 [hep-ph]].
%330 citations counted in INSPIRE as of 21 Jan 2025



%\cite{Machleidt:2011zz}
\bibitem{Machleidt:2011zz}
Machleidt, R.; Entem, D.R.
Chiral effective field theory and nuclear forces.
\emph{Phys. Rep.} \textbf{2011}, \emph{503}, 1--75.
\url{https://doi.org/10.1016/j.physrep.2011.02.001}.
%[arXiv:1105.2919 [nucl-th]].
%1522 citations counted in INSPIRE as of 21 Jan 2025



%\cite{Skyrme:1961vq}
\bibitem{Skyrme:1961vq}
Skyrme, T.H.R.
A Nonlinear field theory.
\emph{Proc. R. Soc. Lond. A} \textbf{1961}, \emph{260}, 127--138.
\url{https://doi.org/10.1098/rspa.1961.0018}.
%2315 citations counted in INSPIRE as of 26 Mar 2023

%\cite{Skyrme:1962vh}
\bibitem{Skyrme:1962vh}
Skyrme, T.H.R.
A Unified Field Theory of Mesons and Baryons.
\emph{Nucl. Phys}. \textbf{1962}, \emph{31}, 556--569.
\url{https://doi.org/10.1016/0029-5582(62)90775-7}.
%1845 citations counted in INSPIRE as of 09 Oct 2023

%\cite{Zahed:1986qz}
\bibitem{Zahed:1986qz}
Zahed, I.; Brown, G.E.
The Skyrme Model.
\emph{Phys. Rep.} \textbf{1986}, \emph{142}, 1--102. %MDPI: please check if the page information is correct.
\url{https://doi.org/10.1016/0370-1573(86)90142-0}.
%687 citations counted in INSPIRE as of 12 Sep 2024

%\cite{Capstick:1986ter}
\bibitem{Capstick:1986ter}
Capstick, S.; Isgur, N.
Baryons in a relativized quark model with chromodynamics.
\emph{Phys. Rev. D} \textbf{1986}, \emph{34}, 2809--2835.
\url{https://doi.org/10.1103/physrevd.34.2809}.
%1498 citations counted in INSPIRE as of 18 Feb 2025

%\cite{Capstick:1986ter,Koniuk:1979vy,Capstick:2000qj}
\bibitem{Koniuk:1979vy}
Koniuk, R.; Isgur, N.
Baryon Decays in a Quark Model with Chromodynamics.
\emph{Phys. Rev. D} \textbf{1980}, \emph{21}, 1868. Erratum in \emph{Phys. Rev. D} \textbf{1981}, \emph{23}, 818.
\url{https://doi.org/10.1103/PhysRevD.21.1868}.
%633 citations counted in INSPIRE as of 18 Feb 2025

%\cite{Capstick:2000qj}
\bibitem{Capstick:2000qj}
Capstick, S.; Roberts, W.
Quark models of baryon masses and decays.
\emph{Prog. Part. Nucl. Phys.} \textbf{2000}, \emph{45}, S241--S331.
\url{https://doi.org/10.1016/S0146-6410(00)00109-5}.
%[arXiv:nucl-th/0008028 [nucl-th]].
%479 citations counted in INSPIRE as of 18 Feb 2025


%\cite{Ma:2016npf}
\bibitem{Ma:2016npf}
Ma, Y.L.; Harada, M.
Lecture notes on the Skyrme model. {\emph{{arXiv} }} \textbf{{2016} %MDPI: we added information, please confirm. please check the whole reference part
}, arXiv:1604.04850.
%[ [hep-ph]].
%18 citations counted in INSPIRE as of 26 Nov 2024



%\cite{tHooft:1973alw}
\bibitem{tHooft:1973alw}
't Hooft, G.
A Planar Diagram Theory for Strong Interactions.
\emph{Nucl. Phys. B} \textbf{1974}, \emph{72}, 461.
\url{https://doi.org/10.1016/0550-3213(74)90154-0}.
%5410 citations counted in INSPIRE as of 26 Mar 2023

%\cite{Witten:1979kh}
\bibitem{Witten:1979kh}
Witten, E.
Baryons in the 1/n Expansion.
\emph{Nucl. Phys. B} \textbf{1979}, \emph{160}, 57--115.
\url{https://doi.org/10.1016/0550-3213(79)90232-3}.
%2860 citations counted in INSPIRE as of 26 Mar 2023


%\cite{Adkins:1983ya}
\bibitem{Adkins:1983ya}
Adkins, G.S.; Nappi, C.R.; Witten, E.
Static Properties of Nucleons in the Skyrme Model.
\emph{Nucl. Phys. B} \textbf{1983}, \emph{228}, 552.
\url{https://doi.org/10.1016/0550-3213(83)90559-X}.
%2100 citations counted in INSPIRE as of 12 Sep 2024


%\cite{Witten:1983tw}
\bibitem{Witten:1983tw}
Witten, E.
Global Aspects of Current Algebra.
\emph{Nucl. Phys. B} \textbf{1983}, \emph{223}, 422--432.
\url{https://doi.org/10.1016/0550-3213(83)90063-9}.
%3100 citations counted in INSPIRE as of 12 Sep 2024



%\cite{Nadkarni:1984eg}
\bibitem{Nadkarni:1984eg}
Nadkarni, S.; Nielsen, H.B.; Zahed, I.
Bosonization Relations as Bag Boundary Conditions.
\emph{Nucl. Phys. B} \textbf{1985}, \emph{253}, 308--322.
\url{https://doi.org/10.1016/0550-3213(85)90533-4}.
%155 citations counted in INSPIRE as of 21 Jan 2025


%\cite{Nadkarni:1985dn}
\bibitem{Nadkarni:1985dn}
Nadkarni, S.; Zahed, I.
Nonabelian Cheshire Cat Bag Models in (1+1)-dimensions.
\emph{Nucl. Phys. B} \textbf{1986}, \emph{263}, 23--36.
\url{https://doi.org/10.1016/0550-3213(86)90025-8}.
%43 citations counted in INSPIRE as of 21 Jan 2025


%\cite{Rho:1993gr}
\bibitem{Rho:1993gr}
Rho, M.
Cheshire cat hadrons.
\emph{Phys. Rep.} \textbf{1994}, \emph{240}, {1--142.} %MDPI: please check if the page information is correct.
\url{https://doi.org/10.1016/0370-1573(94)90002-7}.
%[arXiv:hep-ph/9310300 [hep-ph]].
%20 citations counted in INSPIRE as of 21 Jan 2025


%\cite{Nielsen:1991bk}
\bibitem{Nielsen:1991bk}
Nielsen, H.B.; Rho, M.; Wirzba, A.; Zahed, I.
Color anomaly in a hybrid bag model.
\emph{Phys. Lett. B} \textbf{1991}, \emph{269}, 389--393.
\url{https://doi.org/10.1016/0370-2693(91)90189-W}.
%34 citations counted in INSPIRE as of 21 Jan 2025


%\cite{Rho:1983bh}
\bibitem{Rho:1983bh}
Rho, M.; Goldhaber, A.S.; Brown, G.E.
Topological Soliton Bag Model for Baryons.
\emph{Phys. Rev. Lett.} \textbf{1983}, \emph{51}, 747--750.
\url{https://doi.org/10.1103/PhysRevLett.51.747}.
%271 citations counted in INSPIRE as of 21 Jan 2025



%\cite{Witten:1979vv}
\bibitem{Witten:1979vv}
Witten, E.
Current Algebra Theorems for the U(1) Goldstone Boson.
\emph{Nucl. Phys. B} \textbf{1979}, \emph{156}, 269--283.
\url{https://doi.org/10.1016/0550-3213(79)90031-2}.
%1711 citations counted in INSPIRE as of 29 Aug 2024



%\cite{Veneziano:1979ec}
\bibitem{Veneziano:1979ec}
Veneziano, G.
U(1) Without Instantons.
\emph{Nucl. Phys. B} \textbf{1979}, \emph{159}, 213--224.
\url{https://doi.org/10.1016/0550-3213(79)90332-8}.
%1534 citations counted in INSPIRE as of 29 Aug 2024



%\cite{DiVecchia:1980yfw}
\bibitem{DiVecchia:1980yfw}
Vecchia, P.D.; Veneziano, G.
Chiral Dynamics in the Large n Limit.
\emph{Nucl. Phys. B} \textbf{1980}, \emph{171}, 253--272.
\url{https://doi.org/10.1016/0550-3213(80)90370-3}.
%801 citations counted in INSPIRE as of 13 Sep 2024

%\cite{Ohta:1981ai}
\bibitem{Ohta:1981ai}
Ohta, N.
Vacuum Structure and Chiral Charge Quantization in the Large $N$ Limit.
\emph{Prog. Theor. Phys.} \textbf{1981}, \emph{66}, 1408. Erratum in \emph{Prog. Theor. Phys.} \textbf{1982}, \emph{67}, 993.
\url{https://doi.org/10.1143/PTP.66.1408}.
%90 citations counted in INSPIRE as of 20 Feb 2025


%\cite{Kawarabayashi:1980uh}
\bibitem{Kawarabayashi:1980uh}
Kawarabayashi, K.; Ohta, N.
On the Partial Conservation of the U(1) Current.
\emph{Prog. Theor. Phys.} \textbf{1981}, \emph{66}, 1789.
\url{https://doi.org/10.1143/PTP.66.1789}.
%140 citations counted in INSPIRE as of 20 Feb 2025
%\cite{Kawarabayashi:1980dp}

\bibitem{Kawarabayashi:1980dp}
Kawarabayashi, K.; Ohta, N.
The Problem of $\eta$ in the Large $N$ Limit: Effective Lagrangian Approach.
\emph{Nucl. Phys. B} \textbf{1980}, \emph{175}, 477--492.
\url{https://doi.org/10.1016/0550-3213(80)90024-3}.
%324 citations counted in INSPIRE as of 20 Feb 2025


%\cite{Ma:2020nih}
\bibitem{Ma:2020nih}
Ma, Y.L.; Rho, M.
Dichotomy of Baryons as Quantum Hall Droplets and Skyrmions: Topological Structure of Dense Matter.
\emph{Symmetry} \textbf{2021}, \emph{13}, 1888.
\url{https://doi.org/10.3390/sym13101888}.
%[arXiv:2009.09219 [nucl-th]].
%20 citations counted in INSPIRE as of 21 Jan 2025





%\cite{Kapustin:2014lwa}
\bibitem{Kapustin:2014lwa}
Kapustin, A.; Thorngren, R.
Anomalies of discrete symmetries in three dimensions and group cohomology.
\emph{Phys. Rev. Lett. }\textbf{2014}, \emph{112}, 231602.
\url{https://doi.org/10.1103/PhysRevLett.112.231602}.
%[arXiv:1403.0617 [hep-th]].
%121 citations counted in INSPIRE as of 14 Sep 2024



%\cite{Kapustin:2014gua}
\bibitem{Kapustin:2014gua}
Kapustin, A.; Seiberg, N.
Coupling a QFT to a TQFT and Duality.
\emph{J. High Energy Phys.} \textbf{2014}, \emph{{2014} }, {1}. %MDPI: we revised the volume and page, please confirm. same below.
\url{https://doi.org/10.1007/JHEP04(2014)001}.
%[arXiv:1401.0740 [hep-th]].
%386 citations counted in INSPIRE as of 14 Sep 2024



%\cite{Gaiotto:2014kfa}
\bibitem{Gaiotto:2014kfa}
Gaiotto, D.; Kapustin, A.; Seiberg, N.; Willett, B.
Generalized Global Symmetries.
\emph{J. High Energy Phys.} \textbf{2015}, \emph{{2015} %MDPI: 
}, 172.
\url{https://doi.org/10.1007/JHEP02(2015)172}.
%[arXiv:1412.5148 [hep-th]].
%1202 citations counted in INSPIRE as of 14 Sep 2024


%\cite{Gaiotto:2017yup}
\bibitem{Gaiotto:2017yup}
Gaiotto, D.; Kapustin, A.; Komargodski, Z.; Seiberg, N.
Theta, Time Reversal, and Temperature.
\emph{J. High Energy Phys.} \textbf{2017}, \emph{{2017} %MDPI: 
}, 91.
\url{https://doi.org/10.1007/JHEP05(2017)091}.
%[arXiv:1703.00501 [hep-th]].
%410 citations counted in INSPIRE as of 27 Aug 2024



%\cite{Gaiotto:2017tne}
\bibitem{Gaiotto:2017tne}
Gaiotto, D.; Komargodski, Z.; Seiberg, N.
Time-reversal breaking in QCD$_{4}$, walls, and dualities in 2 + 1 dimensions.
\emph{J. High Energy Phys.} \textbf{2018}, \emph{{2018} %MDPI: 
}, 110.
\url{https://doi.org/10.1007/JHEP01(2018)110}.
%[arXiv:1708.06806 [hep-th]].
%214 citations counted in INSPIRE as of 27 Aug 2024



%\cite{Hsin:2016blu}
\bibitem{Hsin:2016blu}
Hsin, P.S.; Seiberg, N.
Level/rank Duality and Chern-Simons-Matter Theories.
\emph{J. High Energy Phys.} \textbf{2016}, \emph{{2016}}, 95.
\url{https://doi.org/10.1007/JHEP09(2016)095}.
%[arXiv:1607.07457 [hep-th]].
%237 citations counted in INSPIRE as of 14 Sep 2024



%\cite{Komargodski:2018odf}
\bibitem{Komargodski:2018odf}
Komargodski, Z.;
Baryons as Quantum Hall Droplets. \emph{arXiv} \textbf{2018}, arXiv:1812.09253.
%[arXiv:1812.09253 [hep-th]].
%41 citations counted in INSPIRE as of 27 Aug 2024

%\cite{Ma:2019xtx}
\bibitem{Ma:2019xtx}
Ma, Y.L.; Nowak, M.A.; Rho, M.; Zahed, I.
Baryon as a Quantum Hall Droplet and the Cheshire Cat Principle.
\emph{Phys. Rev. Lett.} \textbf{2019}, \emph{123}, 172301.
\url{https://doi.org/10.1103/PhysRevLett.123.172301}.
%[arXiv:1907.00958 [hep-th]].
%23 citations counted in INSPIRE as of 27 Aug 2024

%\cite{Ma:2019ery}
\bibitem{Ma:2019ery}
Ma, Y.L.; Rho, M.
Towards the hadron\textendash{}quark continuity via a topology change in compact stars.
\emph{Prog. Part. Nucl. Phys.} \textbf{2020}, \emph{113}, 103791.
\url{https://doi.org/10.1016/j.ppnp.2020.103791}.
%[arXiv:1909.05889 [nucl-th]].
%59 citations counted in INSPIRE as of 21 Jan 2025

%\cite{Lin:2023qya}
\bibitem{Lin:2023qya}
Lin, F.; Ma, Y.L.
Baryons as vortexes on the \ensuremath{\eta}' domain wall.
\emph{J. High Energy Phys.} \textbf{2024}, \emph{{2014} %MDPI: 
}, 270.
\url{https://doi.org/10.1007/JHEP05(2024)270}.
%[arXiv:2310.16438 [hep-th]].
%1 citations counted in INSPIRE as of 27 Aug 2024


%\cite{Zhang:1988wy}
\bibitem{Zhang:1988wy}
Zhang, S.C.; Hansson, T.H.; Kivelson, S.
An effective field theory model for the fractional quantum hall effect.
\emph{Phys. Rev. Lett.} \textbf{1988}, \emph{62}, 82--85.
\url{https://doi.org/10.1103/PhysRevLett.62.82}.
%666 citations counted in INSPIRE as of 18 Nov 2024


%\cite{Chodos:1974je}
\bibitem{Chodos:1974je}
Chodos, A.; Jaffe, R.L.; Johnson, K.; Thorn, C.B.; Weisskopf, V.F.
A New Extended Model of Hadrons.
\emph{Phys. Rev. D} \textbf{1974}, \emph{9}, 3471--3495.
\url{https://doi.org/10.1103/PhysRevD.9.3471}.
%3289 citations counted in INSPIRE as of 12 Sep 2024

%\cite{Thomas:1982kv}
\bibitem{Thomas:1982kv}
Thomas, A.W.
Chiral Symmetry and the Bag Model: A New Starting Point for Nuclear Physics.
\emph{Adv. Nucl. Phys.} \textbf{1984}, \emph{13}, {1--137.} %MDPI: please check if the page information is correct.
\url{https://doi.org/10.1007/978-1-4613-9892-9\_1}.
%788 citations counted in INSPIRE as of 13 Sep 2024


%\cite{Chodos:1975ix}
\bibitem{Chodos:1975ix}
Chodos, A.; Thorn, C.B.
Chiral Hedgehogs in the Bag Theory.
\emph{Phys. Rev. D} \textbf{1975}, \emph{12}, 2733.
\url{https://doi.org/10.1103/PhysRevD.12.2733}.
%464 citations counted in INSPIRE as of 13 Sep 2024

%\cite{Mulders:1984df}
\bibitem{Mulders:1984df}
Mulders, P.J.
Theoretical Aspects of Hybrid Chiral Bag Models.
\emph{Phys. Rev. D} \textbf{1984}, \emph{30}, 1073.
\url{https://doi.org/10.1103/PhysRevD.30.1073}.
%83 citations counted in INSPIRE as of 13 Sep 2024

%\cite{Goldman:1980ww}
\bibitem{Goldman:1980ww}
Goldman, J.T.; Haymaker, R.W.
Dynamically Broken Chiral Symmetry With Bag Confinement.
\emph{Phys. Rev. D} \textbf{1981}, \emph{24}, 724.
\url{https://doi.org/10.1103/PhysRevD.24.724}.
%95 citations counted in INSPIRE as of 13 Sep 2024


%\cite{Nadkarni:1985dm}
\bibitem{Nadkarni:1985dm}
Nadkarni, S.; Nielsen, H.B.
Partial Bosonization: The Formalism of Cheshire Cat Bag Models.
\emph{Nucl. Phys. B} \textbf{1986}, \emph{263}, 1--22.
\url{https://doi.org/10.1016/0550-3213(86)90024-6}.
%51 citations counted in INSPIRE as of 22 Aug 2024



%\cite{Perry:1986sz}
\bibitem{Perry:1986sz}
Perry, R.J.; Rho, M.
Removing Bag Dynamics From Chiral Bag Models: An Illustrative Example.
\emph{Phys. Rev. D} \textbf{1986}, \emph{34}, 1169--1175.
\url{https://doi.org/10.1103/PhysRevD.34.1169}.
%30 citations counted in INSPIRE as of 13 Sep 2024

%\cite{Damgaard:1992cy}
\bibitem{Damgaard:1992cy}
Damgaard, P.H.; Nielsen, H.B.; Sollacher, R.
Smooth bosonization: The Cheshire cat revisited.
\emph{Nucl. Phys. B} \textbf{1992}, \emph{385}, 227--250.
\url{https://doi.org/10.1016/0550-3213(92)90100-P}.
%76 citations counted in INSPIRE as of 13 Sep 2024



%\cite{Nielsen:1992va}
\bibitem{Nielsen:1992va}
Nielsen, H.B.; Rho, M.; Wirzba, A.; Zahed, I.
The tale of the eta-prime from the cheshire cat principle.
\emph{Phys. Lett. B} \textbf{1992}, \emph{281}, 345--350.
\url{https://doi.org/10.1016/0370-2693(92)91153-Z}.
%24 citations counted in INSPIRE as of 26 Aug 2024


%\cite{Lin:2025lzf}
\bibitem{Lin:2025lzf}
Lin, F.; Ma, Y.L.
Confined Monopoles in Chiral Bag. \emph{arXiv} \textbf{2025}, arXiv:2501.16653.
%[ [hep-ph]].
%0 citations counted in INSPIRE as of 30 Jan 2025


%\cite{Wess:1971yu}
\bibitem{Wess:1971yu}
Wess, J.; Zumino, B.
Consequences of anomalous Ward identities.
\emph{Phys. Lett. B} \textbf{1971}, \emph{37}, 95--97.
\url{https://doi.org/10.1016/0370-2693(71)90582-X}.
%3089 citations counted in INSPIRE as of 31 Oct 2024


%\cite{Karasik:2022tmd}
\bibitem{Karasik:2022tmd}
Karasik, A.
From Skyrmions to One Flavored Baryons and Beyond.
\emph{Symmetry} \textbf{2022}, \emph{14}, 2347.
\url{https://doi.org/10.3390/sym14112347}.
%1 citations counted in INSPIRE as of 31 Oct 2024


%\cite{Niemi:1984vz}
\bibitem{Niemi:1984vz}
Niemi, A.J.; Semenoff, G.W.
Fermion Number Fractionization in Quantum Field Theory.
\emph{Phys. Rep.} \textbf{1986}, \emph{135}, 99.
\url{https://doi.org/10.1016/0370-1573(86)90167-5}.
%332 citations counted in INSPIRE as of 27 Aug 2024


%\cite{Battye:2006na}
\bibitem{Battye:2006na}
Battye, R.; Manton, N.S.; Sutcliffe, P.
Skyrmions and the alpha-particle model of nuclei.
\emph{Proc. R. Soc. Lond. A} \textbf{2007}, \emph{463}, 261--279.
\url{https://doi.org/10.1098/rspa.2006.1767}.
%[arXiv:hep-th/0605284 [hep-th]].
%97 citations counted in INSPIRE as of 21 Jan 2025


%\cite{Park:2009bb}
\bibitem{Park:2009bb}
Park, B.Y.; Vento, V.
Skyrmion approach to finite density and temperature. \emph{arXiv} \textbf{2009}, arXiv:0906.3263.
\url{https://doi.org/10.1142/9789814280709\_0005}.
%[arXiv:0906.3263 [hep-ph]].
%19 citations counted in INSPIRE as of 22 Jan 2025


%\cite{Brown:2010api}
\bibitem{Brown:2010api}
Brown, G.E.; Rho, M.
\emph{The Multifaceted Skyrmion};
World Scientific:  {Singapore,} %MDPI: We added the location of publisher. Please confirm
 2010; ISBN 978-981-4280-69-3; 978-981-4467-16-2.
\url{https://doi.org/10.1142/7397}.
%3 citations counted in INSPIRE as of 21 Jan 2025


%\cite{Ma:2016gdd}
\bibitem{Ma:2016gdd}
Ma, Y.L.; Rho, M.
Recent progress on dense nuclear matter in skyrmion approaches.
\emph{Sci. China Phys. Mech. Astron.} \textbf{2017}, \emph{60}, 032001.
\url{https://doi.org/10.1007/s11433-016-0497-2}.
%[arXiv:1612.06600 [nucl-th]].
%30 citations counted in INSPIRE as of 21 Jan 2025

%\cite{Naya:2018kyi}
\bibitem{Naya:2018kyi}
Naya, C.; Sutcliffe, P.
Skyrmions and clustering in light nuclei.
\emph{Phys. Rev. Lett.} \textbf{2018}, \emph{121}, 232002.
\url{https://doi.org/10.1103/PhysRevLett.121.232002}.
%[arXiv:1811.02064 [hep-th]].
%59 citations counted in INSPIRE as of 21 Jan 2025


%\cite{Manton:2022fcb}
\bibitem{Manton:2022fcb}
Manton, N.S.
\emph{Skyrmions---A Theory of Nuclei};
World Scientific:  {Singapore,} %MDPI: We added the location of publisher. Please confirm.
2022; ISBN 978-1-80061-247-1; 978-1-80061-249-5.
\url{https://doi.org/10.1142/q0368}.
%20 citations counted in INSPIRE as of 21 Jan 2025


%\cite{Wang:2023qxq}
\bibitem{Wang:2023qxq}
Wang, J.S.; Ma, Y.L.
Vector meson effects on multi-Skyrmion states from the rational map ansatz.
\emph{Sci. China Phys. Mech. Astron.} \textbf{2023}, \emph{66}, 112011.
\url{https://doi.org/10.1007/s11433-023-2220-y}.
%[arXiv:2305.17357 [hep-ph]].
%0 citations counted in INSPIRE as of 21 Jan 2025


%\cite{Kugler:1988mu}
\bibitem{Kugler:1988mu}
Kugler, M.; Shtrikman, S.
A New Skyrmion Crystal.
\emph{Phys. Lett. B} \textbf{1988}, \emph{208}, 491--494.
\url{https://doi.org/10.1016/0370-2693(88)90653-3}.
%114 citations counted in INSPIRE as of 22 Jan 2025


%\cite{Kugler:1989uc}
\bibitem{Kugler:1989uc}
Kugler, M.; Shtrikman, S.
Skyrmion Crystals and Their Symmetries.
\emph{Phys. Rev. D} \textbf{1989}, \emph{40}, 3421.
\url{https://doi.org/10.1103/PhysRevD.40.3421}.
%98 citations counted in INSPIRE as of 22 Jan 2025




%\cite{Goldhaber:1987pb}
\bibitem{Goldhaber:1987pb}
Goldhaber, A.S.; Manton, N.S.
Maximal Symmetry of the Skyrme Crystal.
\emph{Phys. Lett. B} \textbf{1987}, \emph{198}, 231--234.
\url{https://doi.org/10.1016/0370-2693(87)91502-4}.
%117 citations counted in INSPIRE as of 22 Jan 2025


%\cite{Lee:2003eg}
\bibitem{Lee:2003eg}
Lee, H.J.; Park, B.Y.; Rho, M.; Vento, V.
Sliding vacua in dense skyrmion matter.
\emph{Nucl. Phys. A} \textbf{2003}, \emph{726}, 69--92.
\url{https://doi.org/10.1016/S0375-9474(03)01626-9}.
%[arXiv:hep-ph/0304066 [hep-ph]].
%44 citations counted in INSPIRE as of 22 Jan 2025


%\cite{Lee:2010sw}
\bibitem{Lee:2010sw}
Lee, H.K.; Park, B.Y.; Rho, M.
Half-Skyrmions, Tensor Forces and Symmetry Energy in Cold Dense Matter.
\emph{Phys. Rev. C} \textbf{2011}, \emph{83}, 025206. Erratum in \emph{Phys. Rev. C} \textbf{2011}, \emph{84}, 059902.
\url{https://doi.org/10.1103/PhysRevC.84.059902}.
%[arXiv:1005.0255 [nucl-th]].
%83 citations counted in INSPIRE as of 26 Jan 2025


%\cite{Holt:2007ih}
\bibitem{Holt:2007ih}
Holt, J.W.; Brown, G.E.; Kuo, T.T.S.; Holt, J.D.; Machleidt, R.
Shell model description of the C-14 dating beta decay with Brown-Rho-scaled NN interactions.
\emph{Phys. Rev. Lett.} \textbf{2008}, \emph{100}, 062501.
\url{https://doi.org/10.1103/PhysRevLett.100.062501}.
%[arXiv:0710.0310 [nucl-th]].
%69 citations counted in INSPIRE as of 26 Jan 2025



%\cite{Paeng:2015noa}
\bibitem{Paeng:2015noa}
Paeng, W.G.; Kuo, T.T.S.; Lee, H.K.; Rho, M.
Scale-Invariant Hidden Local Symmetry, Topology Change and Dense Baryonic Matter.
\emph{Phys. Rev. C} \textbf{2016}, \emph{93}, 055203.
\url{https://doi.org/10.1103/PhysRevC.93.055203}.
%[arXiv:1508.05210 [hep-ph]].
%43 citations counted in INSPIRE as of 26 Jan 2025



%\cite{Paeng:2017qvp}
\bibitem{Paeng:2017qvp}
Paeng, W.G.; Kuo, T.T.S.; Lee, H.K.; Ma, Y.L.; Rho, M.
Scale-invariant hidden local symmetry, topology change, and dense baryonic matter. II.
\emph{Phys. Rev. D} \textbf{2017}, \emph{96}, 014031.
\url{https://doi.org/10.1103/PhysRevD.96.014031}.
%[arXiv:1704.02775 [nucl-th]].
%59 citations counted in INSPIRE as of 26 Jan 2025


\bibitem{Ma:2018jze}
Ma, Y.L.; Lee, H.K.; Paeng, W.G.; Rho, M.
Pseudoconformal equation of state in compact-star matter from topology change and hidden symmetries of QCD.
\emph{Sci. China Phys. Mech. Astron.} \textbf{2019}, \emph{62}, 112011.
\url{https://doi.org/10.1007/s11433-019-9399-5}.
%[arXiv:1804.00305 [nucl-th]].
%16 citations counted in INSPIRE as of 26 Jan 2025


%\cite{Hosaka:1996ee}
\bibitem{Hosaka:1996ee}
Hosaka, A.; Toki, H.
Chiral bag model for the nucleon.
\emph{Phys. Rep.} \textbf{1996}, \emph{277}, 65--188.
\url{https://doi.org/10.1016/S0370-1573(96)00013-0}.
%101 citations counted in INSPIRE as of 28 Jan 2025


%\cite{Goldstone:1981kk}
\bibitem{Goldstone:1981kk}
Goldstone, J.; Wilczek, F.
Fractional Quantum Numbers on Solitons.
\emph{Phys. Rev. Lett.} \textbf{1981}, \emph{47}, 986--989.
\url{https://doi.org/10.1103/PhysRevLett.47.986}.
%677 citations counted in INSPIRE as of 13 Sep 2024




%\cite{Jaroszewicz:1983wd}
\bibitem{Jaroszewicz:1983wd}
Jaroszewicz, T.
On the Vacuum Baryon Number in the Chiral Bag Model.
\emph{Phys. Lett. B} \textbf{1984}, \emph{143}, 217--221.
\url{https://doi.org/10.1016/0370-2693(84)90838-4}.
%6 citations counted in INSPIRE as of 19 Nov 2024


%\cite{Rho:2021aad}
\bibitem{Rho:2021aad}
Rho, M.
Skyrmions and Fractional Quantum Hall Droplets Unified by Hidden Symmetries in Dense Matter. \emph{arXiv} \textbf{2021}, arXiv:2109.10059.
%[arXiv:2109.10059 [nucl-th]].
%1 citations counted in INSPIRE as of 26 Jan 2025



%\cite{Karasik:2020zyo}
\bibitem{Karasik:2020zyo}
Karasik, A.
Vector dominance, one flavored baryons, and QCD domain walls from the ``hidden'' Wess-Zumino term.
\emph{SciPost Phys.} \textbf{2021}, \emph{10}, 138.
\url{https://doi.org/10.21468/SciPostPhys.10.6.138}.
%[arXiv:2010.10544 [hep-th]].
%19 citations counted in INSPIRE as of 11 Sep 2024


%\cite{Karasik:2020pwu}
\bibitem{Karasik:2020pwu}
Karasik, A.
Skyrmions, Quantum Hall Droplets, and one current to rule them all.
\emph{SciPost Phys.} \textbf{2020}, \emph{9}, 8.
\url{https://doi.org/10.21468/SciPostPhys.9.1.008}.
%[arXiv:2003.07893 [hep-th]].
%21 citations counted in INSPIRE as of 11 Sep 2024


%\cite{tHooft:1976rip}
\bibitem{tHooft:1976rip}
't Hooft, G.
Symmetry Breaking Through Bell-Jackiw Anomalies.
\emph{Phys. Rev. Lett.} \textbf{1976}, \emph{37}, 8--11.
\url{https://doi.org/10.1103/PhysRevLett.37.8}.
%4276 citations counted in INSPIRE as of 23 Feb 2025


%\cite{tHooft:1986ooh}
\bibitem{tHooft:1986ooh}
't Hooft, G.
How Instantons Solve the U(1) Problem.
\emph{Phys. Rep.} \textbf{1986}, \emph{142}, 357--387.
\url{https://doi.org/10.1016/0370-1573(86)90117-1}.
%819 citations counted in INSPIRE as of 23 Feb 2025



%\cite{Naculich:2007nc}
\bibitem{Naculich:2007nc}
Naculich, S.G.; Schnitzer, H.J.
Level-rank duality of the U(N) WZW model, Chern-Simons theory, and 2-D qYM theory.
\emph{J. High Energy Phys.} \textbf{2007}, \emph{{2007} %MDPI: 
}, 23.
\url{https://doi.org/10.1088/1126-6708/2007/06/023}.
[arXiv:hep-th/0703089 [hep-th]].
%50 citations counted in INSPIRE as of 20 Feb 2025


%\cite{Nakanishi:1990hj}
\bibitem{Nakanishi:1990hj}
Nakanishi, T.; Tsuchiya, A.
Level rank duality of WZW models in conformal field theory.
\emph{Commun. Math. Phys.} \textbf{1992}, \emph{144}, 351--372.
\url{https://doi.org/10.1007/BF02101097}.
%72 citations counted in INSPIRE as of 20 Feb 2025


\bibitem{Aharony:2015mjs}
Aharony, O.
Baryons, monopoles and dualities in Chern-Simons-matter theories.
\emph{J. High Energy Phys.} \textbf{2016}, \emph{{2016} %MDPI: 
}, 93.
\url{https://doi.org/10.1007/JHEP02(2016)093}.
%[arXiv:1512.00161 [hep-th]].
%214 citations counted in INSPIRE as of 20 Feb 2025


%\cite{Naculich:1990pa}
\bibitem{Naculich:1990pa}
Naculich, S.G.; Riggs, H.A.; Schnitzer, H.J.
Group Level Duality in {WZW} Models and {Chern-Simons} Theory.
\emph{Phys. Lett. B} \textbf{1990}, \emph{246}, 417--422.
\url{https://doi.org/10.1016/0370-2693(90)90623-E}.
%108 citations counted in INSPIRE as of 20 Feb 2025

%\cite{Camperi:1990dk}
\bibitem{Camperi:1990dk}
Camperi, M.; Levstein, F.; Zemba, G.
The Large $N$ Limit of {Chern-Simons} Gauge Theory.
\emph{Phys. Lett. B} \textbf{1990}, \emph{247}, 549--554.
\url{https://doi.org/10.1016/0370-2693(90)91899-M}.
%39 citations counted in INSPIRE as of 20 Feb 2025
\url{https://doi.org/10.1016/0550-3213(91)90110-J}.


%\cite{Zahed:1984jmf}
\bibitem{Zahed:1984jmf}
Zahed, I.
Anomalous Baryon Number in (1+1)-Dimensions.
\emph{Phys. Rev. D} \textbf{1984}, \emph{30}, 2221--2226.
\url{https://doi.org/10.1103/PhysRevD.30.2221}.
%27 citations counted in INSPIRE as of 14 Nov 2024


\bibitem{Callan:1984sa}
Callan, C.G.; Jr; Harvey, J.A.
Anomalies and Fermion Zero Modes on Strings and Domain Walls.
\emph{Nucl. Phys. B} \textbf{1985}, \emph{250}, 427--436.
\url{https://doi.org/10.1016/0550-3213(85)90489-4}.
%770 citations counted in INSPIRE as of 09 Oct 2023


%\cite{Jatkar:1989sc}
\bibitem{Jatkar:1989sc}
Jatkar, D.P.; Khare, A.
Peculiar Charged Vortices in Higgs Models With Pure {Chern-Simons} Term.
\emph{Phys. Lett. B} \textbf{1990}, \emph{236}, 283--286.
\url{https://doi.org/10.1016/0370-2693(90)90983-D}.
%40 citations counted in INSPIRE as of 18 Nov 2024

%\cite{Wilczek:1981du}
\bibitem{Wilczek:1981du}
Wilczek, F.
Magnetic Flux, Angular Momentum, and Statistics.
\emph{Phys. Rev. Lett.} \textbf{1982}, \emph{48}, 1144--1146.
\url{https://doi.org/10.1103/PhysRevLett.48.1144}.
%951 citations counted in INSPIRE as of 18 Nov 2024

%\cite{Rao:1992aj}
\bibitem{Rao:1992aj}
Rao, S.
An Anyon primer. \emph{arXiv} \textbf{1992}, arXiv:hep-th/9209066.
%[arXiv:hep-th/9209066 [hep-th]].
%38 citations counted in INSPIRE as of 18 Nov 2024


%\cite{Weigel:1986zc}
\bibitem{Weigel:1986zc}
Weigel, H.; Schwesinger, B.; Holzwarth, G.
Exotic Baryon Number $B=2$ States in the SU(2) Skyrme Model.
\emph{Phys. Lett. B} \textbf{1986}, \emph{168}, 321--325.
\url{https://doi.org/10.1016/0370-2693(86)91637-0}.
%65 citations counted in INSPIRE as of 27 Jan 2025

%\cite{Houghton:1997kg}
\bibitem{Houghton:1997kg}
Houghton, C.J.; Manton, N.S.; Sutcliffe, P.M.
Rational maps, monopoles and Skyrmions.
\emph{Nucl. Phys. B} \textbf{1998}, \emph{510}, 507--537.
\url{https://doi.org/10.1016/S0550-3213(97)00619-6}.
%[arXiv:hep-th/9705151 [hep-th]].
%327 citations counted in INSPIRE as of 27 Jan 2025



%\cite{Peskin:1977kp}
\bibitem{Peskin:1977kp}
Peskin, M.E.
Mandelstam 't Hooft Duality in Abelian Lattice Models.
\emph{Ann. Phys.} \textbf{1978}, \emph{113}, 122.
\url{https://doi.org/10.1016/0003-4916(78)90252-X}.
%435 citations counted in INSPIRE as of 18 Nov 2024



%\cite{Tong:2016kpv}
\bibitem{Tong:2016kpv}
Tong, D.
Lectures on the Quantum Hall Effect. \emph{arXiv} \textbf{2016}, arXiv:1606.06687.
%[arXiv:1606.06687 [hep-th]].
%255 citations counted in INSPIRE as of 18 Nov 2024


%\cite{Eto:2023tuu}
\bibitem{Eto:2023tuu}
M.~Eto, K.~Nishimura and M.~Nitta,
Domain-wall Skyrmion phase in a rapidly rotating QCD matter.
\emph{J. High Energy Phys.} \textbf{2024}, \emph{{2024} %MDPI: 
}, 19.
\url{https://doi.org/10.1007/JHEP03(2024)019}.
%[arXiv:2310.17511 [hep-ph]].
%8 citations counted in INSPIRE as of 12 Apr 2024

%\cite{Eto:2023rzd}
\bibitem{Eto:2023rzd}
Eto, M.; Nishimura, K.; Nitta, M.
Non-Abelian chiral soliton lattice in rotating QCD matter: Nambu-Goldstone and excited modes.
\emph{J. High Energy Phys.} \textbf{2024}, \emph{{2024} %MDPI: 
}, 35.
\url{https://doi.org/10.1007/JHEP03(2024)035}.
%[arXiv:2312.10927 [hep-ph]].
%2 citations counted in INSPIRE as of 13 Apr 2024


%\cite{Kudryavtsev:1996er}
\bibitem{Kudryavtsev:1996er}
Kudryavtsev, A.E.; Piette, B.; Zakrzewski, W.J.
Mesons, baryons and waves in the baby Skyrmion model.
\emph{Eur. Phys. J. C} \textbf{1998}, \emph{1}, 333--341
\url{https://doi.org/10.1007/BF01245822}.
%[arXiv:hep-th/9611217 [hep-th]].
%21 citations counted in INSPIRE as of 27 Jan 2025


%\cite{Battye:2013tka}
\bibitem{Battye:2013tka}
Battye, R.A.; Haberichter, M.
Isospinning baby Skyrmion solutions.
\emph{Phys. Rev. D} \textbf{2013}, \emph{88}, 125016.
\url{https://doi.org/10.1103/PhysRevD.88.125016}.
%[arXiv:1309.3907 [hep-th]].
%29 citations counted in INSPIRE as of 27 Jan 2025


%\cite{Winyard:2015dba}
\bibitem{Winyard:2015dba}
Winyard, T.
Skyrmion and Baby Skyrmion Formation from Domain Walls. \emph{arXiv} \textbf{2015}, arXiv:1507.07482.
%[arXiv:1507.07482 [hep-th]].
%1 citations counted in INSPIRE as of 27 Jan 2025

%\cite{Leask:2024ith}
\bibitem{Leask:2024ith}
Leask, P.
Baby skyrmion crystals stabilized by vector mesons.
\emph{Phys. Lett. B} \textbf{2024}, \emph{855}, 138842.
\url{https://doi.org/10.1016/j.physletb.2024.138842}.
%[arXiv:2403.14810 [hep-th]].
%1 citations counted in INSPIRE as of 27 Jan 2025




%\cite{Bigazzi:2022luo}
\bibitem{Bigazzi:2022luo}
Bigazzi, F.; Cotrone, A.L.; Olzi, A.
Hall Droplet Sheets in Holographic QCD.
\emph{J. High Energy Phys.} \textbf{2023}, \emph{{2023} %MDPI: 
}, 194.
\url{https://doi.org/10.1007/JHEP02(2023)194}.
%[arXiv:2211.05147 [hep-th]].
%3 citations counted in INSPIRE as of 21 Jun 2023


%\cite{Huang:2017pqe}
\bibitem{Huang:2017pqe}
Huang, X.G.; Nishimura, K.; Yamamoto, N.
Anomalous effects of dense matter under rotation.
\emph{J. High Energy Phys.} \textbf{2018}, {2018} %MDPI: 
, 69.
\url{https://doi.org/10.1007/JHEP02(2018)069}.
%[arXiv:1711.02190 [hep-ph]].
%59 citations counted in INSPIRE as of 14 Apr 2024

%\cite{Huang:2019rkz}
\bibitem{Huang:2019rkz}
Huang, X.G.; Nishimura, K.; Yamamoto, N.
Anomaly-Induced Effects of Rotating Dense Matter.
\emph{JPS Conf. Proc.} \textbf{2019}, \emph{26}, 031020.
\url{https://doi.org/10.7566/JPSCP.26.031020}.
%0 citations counted in INSPIRE as of 14 Apr 2024

%\cite{Nishimura:2020odq}
\bibitem{Nishimura:2020odq}
Nishimura, K.; Yamamoto, N.
Topological term, QCD anomaly, and the $\eta^{'}$ chiral soliton lattice in rotating baryonic matter.
\emph{J. High Energy Phys.} \textbf{2020}, \emph{{2020} %MDPI: 
}, 196.
\url{https://doi.org/10.1007/JHEP07(2020)196}.
%[arXiv:2003.13945 [hep-ph]].
%22 citations counted in INSPIRE as of 14 Apr 2024

%\cite{Eto:2021gyy}
\bibitem{Eto:2021gyy}
Eto, M.; Nishimura, K.; Nitta, M.
Phases of rotating baryonic matter: non-Abelian chiral soliton lattices, antiferro-isospin chains, and ferri/ferromagnetic magnetization.
\emph{J. High Energy Phys.} \textbf{2022}, \emph{{2022} %MDPI: 
}, 305.
\url{https://doi.org/10.1007/JHEP08(2022)305}.
%[arXiv:2112.01381 [hep-ph]].
%17 citations counted in INSPIRE as of 14 Apr 2024

%\cite{Nitta:2012wi}
\bibitem{Nitta:2012wi}
Nitta, M.
Correspondence between Skyrmions in 2+1 and 3+1 Dimensions.
\emph{Phys. Rev. D} \textbf{2013}, \emph{87}, 025013.
\url{https://doi.org/10.1103/PhysRevD.87.025013}.
%[arXiv:1210.2233 [hep-th]].
%58 citations counted in INSPIRE as of 14 Apr 2024

%\cite{Eto:2015uqa}
\bibitem{Eto:2015uqa}
Eto, M.; Nitta, M.
Non-Abelian Sine-Gordon Solitons: Correspondence between $SU(N)$ Skyrmions and ${\mathbb C}P^{N-1}$ Lumps.
\emph{Phys. Rev. D} \textbf{2015}, \emph{91}, 085044.
\url{https://doi.org/10.1103/PhysRevD.91.085044}.
%[arXiv:1501.07038 [hep-th]].
%23 citations counted in INSPIRE as of 14 Apr 2024

%\cite{Nitta:2022ahj}
\bibitem{Nitta:2022ahj}
Nitta, M.
Relations among topological solitons.
\emph{Phys. Rev. D} \textbf{2022}, \emph{105}, 105006.
\url{https://doi.org/10.1103/PhysRevD.105.105006}.
%[arXiv:2202.03929 [hep-th]].
%12 citations counted in INSPIRE as of 14 Apr 2024


%\cite{Gudnason:2014nba}
\bibitem{Gudnason:2014nba}
Gudnason, S.B.; Nitta, M.
Domain wall Skyrmions.
\emph{Phys. Rev. D} \textbf{2014}, \emph{89}, 085022.
\url{https://doi.org/10.1103/PhysRevD.89.085022}.
%[arXiv:1403.1245 [hep-th]].
%45 citations counted in INSPIRE as of 14 Apr 2024


%\cite{Eto:2023lyo}
\bibitem{Eto:2023lyo}
Eto, M.; Nishimura, K.; Nitta, M.
How baryons appear in low-energy QCD: Domain-wall Skyrmion phase in strong magnetic fields. \emph{arXiv} \textbf{2023}, arXiv:2304.02940.
%[arXiv:2304.02940 [hep-ph]].
%11 citations counted in INSPIRE as of 12 Apr 2024


%\cite{Eto:2023wul}
\bibitem{Eto:2023wul}
Eto, M.; Nishimura, K.; Nitta, M.
Phase diagram of QCD matter with magnetic field: domain-wall Skyrmion chain in chiral soliton lattice.
\emph{J. High Energy Phys.} \textbf{2023}, \emph{{2023} %MDPI: 
}, 32.
\url{https://doi.org/10.1007/JHEP12(2023)032}.
%[arXiv:2311.01112 [hep-ph]].
%7 citations counted in INSPIRE as of 12 Apr 2024

%\cite{Qiu:2024zpg}
\bibitem{Qiu:2024zpg}
Qiu, Z.; Nitta, M.
Baryonic Vortex Phase and Magnetic Field Generation in QCD with Isospin and Baryon Chemical Potentials. \emph{arXiv} \textbf{2024}, arXiv:2403.07433.
%[arXiv:2403.07433 [hep-ph]].
%0 citations counted in INSPIRE as of 14 Apr 2024

%\cite{Chen:2021vou}
\bibitem{Chen:2021vou}
Chen, S.; Fukushima, K.; Qiu, Z.
Skyrmions in a magnetic field and \ensuremath{\pi}0 domain wall formation in dense nuclear matter.
\emph{Phys. Rev. D} \textbf{2022}, \emph{105}, L011502.
\url{https://doi.org/10.1103/PhysRevD.105.L011502}.
%[arXiv:2104.11482 [hep-ph]].
%15 citations counted in INSPIRE as of 14 Apr 2024

%\cite{Fukushima:2018ohd}
\bibitem{Fukushima:2018ohd}
Fukushima, K.; Imaki, S.
Anomaly inflow on QCD axial domain-walls and vortices.
\emph{Phys. Rev. D} \textbf{2018}, \emph{97}, 114003.
\url{https://doi.org/10.1103/PhysRevD.97.114003}.
%[arXiv:1802.08096 [hep-ph]].
%6 citations counted in INSPIRE as of 14 Apr 2024


%\cite{Dierigl:2014xta}
\bibitem{Dierigl:2014xta}
Dierigl, M.; Pritzel, A.
Topological Model for Domain Walls in (Super-)Yang-Mills Theories.
\emph{Phys. Rev. D} \textbf{2014}, \emph{90}, 105008.
\url{https://doi.org/10.1103/PhysRevD.90.105008}.
%[arXiv:1405.4291 [hep-th]].
%32 citations counted in INSPIRE as of 28 Aug 2024


%\cite{Banks:2010zn}
\bibitem{Banks:2010zn}
Banks, T.; Seiberg, N.
Symmetries and Strings in Field Theory and Gravity.
\emph{Phys. Rev. D} \textbf{2011}, \emph{83}, 084019.
\url{https://doi.org/10.1103/PhysRevD.83.084019}.
%[arXiv:1011.5120 [hep-th]].
%854 citations counted in INSPIRE as of 28 Aug 2024

%\cite{Gukov:2013zka}
\bibitem{Gukov:2013zka}
Gukov, S.; Kapustin, A.
Topological Quantum Field Theory, Nonlocal Operators, and Gapped Phases of Gauge Theories. \emph{arXiv} \textbf{2013}, arXiv:1307.4793.
%[arXiv:1307.4793 [hep-th]].
%91 citations counted in INSPIRE as of 28 Aug 2024

%\cite{Horowitz:1989ng}
\bibitem{Horowitz:1989ng}
Horowitz, G.T.
Exactly Soluble Diffeomorphism Invariant Theories.
\emph{Commun. Math. Phys.} \textbf{1989}, \emph{125}, 417.
\url{https://doi.org/10.1007/BF01218410}.
%459 citations counted in INSPIRE as of 28 Aug 2024


%\cite{Kitano:2020evx}
\bibitem{Kitano:2020evx}
Kitano, R.; Matsudo, R.
Vector mesons on the wall.
\emph{J. High Energy Phys.} \textbf{2021}, \emph{{2021} %MDPI: 
}, 23.
\url{https://doi.org/10.1007/JHEP03(2021)023}.
%[arXiv:2011.14637 [hep-th]].
%12 citations counted in INSPIRE as of 29 Aug 2024

%\cite{Witten:1979ey}
\bibitem{Witten:1979ey}
Witten, E.
Dyons of Charge e theta/2 pi.
\emph{Phys. Lett. B} \textbf{1979}, \emph{86}, 283--287/
\url{https://doi.org/10.1016/0370-2693(79)90838-4}.
%938 citations counted in INSPIRE as of 27 Jan 2025


%\cite{Harada:1992np}
\bibitem{Harada:1992np}
Harada, M.; Yamawaki, K.
Hidden local symmetry at one loop.
\emph{Phys. Lett. B} \textbf{1992}, \emph{297}, 151--158.
\url{https://doi.org/10.1016/0370-2693(92)91084-M}.
%[arXiv:hep-ph/9210208 [hep-ph]].
%75 citations counted in INSPIRE as of 27 Jan 2025


%\cite{Harada:2003jx}
\bibitem{Harada:2003jx}
Harada, M.; Yamawaki, K.
Hidden local symmetry at loop: A New perspective of composite gauge boson and chiral phase transition.
\emph{Phys. Rep.} \textbf{2003}, \emph{381}, {1--233.} %MDPI: please check if the page information is correct.
\url{https://doi.org/10.1016/S0370-1573(03)00139-X}.
%[arXiv:hep-ph/0302103 [hep-ph]].
%646 citations counted in INSPIRE as of 27 Jan 2025

%\cite{Bando:1984ej}
\bibitem{Bando:1984ej}
Bando, M.; Kugo, T.; Uehara, S.; Yamawaki, K.; Yanagida, T.
Is rho Meson a Dynamical Gauge Boson of Hidden Local Symmetry?
\emph{Phys. Rev. Lett.} \textbf{1985}, \emph{54}, 1215.
\url{https://doi.org/10.1103/PhysRevLett.54.1215}.
%1030 citations counted in INSPIRE as of 27 Jan 2025


%\cite{Bando:1987br}
\bibitem{Bando:1987br}
Bando, M.; Kugo, T.; Yamawaki, K.
Nonlinear Realization and Hidden Local Symmetries.
\emph{Phys. Rep.} \textbf{1988}, \emph{164}, 217--314.
\url{https://doi.org/10.1016/0370-1573(88)90019-1}.
%1431 citations counted in INSPIRE as of 27 Jan 2025



%\cite{Goldstone:1983tu}
\bibitem{Goldstone:1983tu}
Goldstone, J.; Jaffe, R.L.
The Baryon Number in Chiral Bag Models.
\emph{Phys. Rev. Lett.} \textbf{1983}, \emph{51}, 1518.
\url{https://doi.org/10.1103/PhysRevLett.51.1518}.
%236 citations counted in INSPIRE as of 27 Aug 2024

%\cite{Dreiner:1988jg}
\bibitem{Dreiner:1988jg}
Dreiner, H.K.; Ellis, J.R.; Flores, R.A.
The Spin of the Proton in a Hybrid Chiral Bag Model.
\emph{Phys. Lett. B} \textbf{1989}, \emph{221}, 167--172.
\url{https://doi.org/10.1016/0370-2693(89)91492-5}.
%34 citations counted in INSPIRE as of 13 Sep 2024

%\cite{Crewther:2013vea}
\bibitem{Crewther:2013vea}
Crewther, R.J.; Tunstall, L.C.
$\Delta I=1/2$ rule for kaon decays derived from QCD infrared fixed point.
\emph{Phys. Rev. D} \textbf{2015}, \emph{91}, 034016.
\url{https://doi.org/10.1103/PhysRevD.91.034016}.
%[arXiv:1312.3319 [hep-ph]].
%101 citations counted in INSPIRE as of 23 Feb 2025

%\cite{Li:2016uzn}
\bibitem{Li:2016uzn}
Li, Y.L.; Ma, Y.L.; Rho, M.
Chiral-scale effective theory including a dilatonic meson.
\emph{Phys. Rev. D} \textbf{2017}, \emph{95}, 114011.
\url{https://doi.org/10.1103/PhysRevD.95.114011}.
%[arXiv:1609.07014 [hep-ph]].
%45 citations counted in INSPIRE as of 23 Feb 2025


%\cite{Harada:2000kb}
\bibitem{Harada:2000kb}
Harada, M.; Yamawaki, K.
Vector manifestation of the chiral symmetry.
\emph{Phys. Rev. Lett.} \textbf{2001}, \emph{86}, 757--760.
\url{https://doi.org/10.1103/PhysRevLett.86.757}.
%[arXiv:hep-ph/0010207 [hep-ph]].
%114 citations counted in INSPIRE as of 30 Jan 2025

%\cite{Harada:2001rf}
\bibitem{Harada:2001rf}
Harada, M.; Yamawaki, K.
Fate of vector dominance in the effective field theory.
\emph{Phys. Rev. Lett.} \textbf{2001}, \emph{87}, 152001.
\url{https://doi.org/10.1103/PhysRevLett.87.152001}.
%[arXiv:hep-ph/0105335 [hep-ph]].
%36 citations counted in INSPIRE as of 30 Jan 2025

\newpage
%\cite{Georgi:1989xy}
\bibitem{Georgi:1989xy}
Georgi, H.
Vector Realization of Chiral Symmetry.
\emph{Nucl. Phys. B} \textbf{1990}, \emph{331}, 311--330.
\url{https://doi.org/10.1016/0550-3213(90)90210-5}.
%168 citations counted in INSPIRE as of 30 Jan 2025


%\cite{Ma:2023ugl}
\bibitem{Ma:2023ugl}
Ma, Y.L.; Yang, W.C.
Topology and Emergent Symmetries in Dense Compact Star Matter.
\emph{Symmetry} \textbf{2023}, \emph{15}, 776.
\url{https://doi.org/10.3390/sym15030776}.
%[arXiv:2301.02105 [nucl-th]].
%6 citations counted in INSPIRE as of 23 Feb 2025

\end{thebibliography}
\end{document}